\documentclass[12pt,a4paper]{article}

\usepackage[left=1in, right=1in, top=1in, bottom=1in, twoside=false]{geometry}
\usepackage[onehalfspacing]{setspace}
\usepackage[]{natbib}

\usepackage{amsmath, amsthm, amsfonts, amssymb}
\usepackage{enumerate}

\usepackage{comment}
\usepackage{caption}
\usepackage{subcaption}
\usepackage{multicol}
\usepackage{bbm}
\usepackage[colorlinks=true,linkcolor=blue, citecolor=blue]{hyperref}

\usepackage[dvips]{graphicx}
\usepackage[dvipsnames]{xcolor}
\usepackage[normalem]{ulem}
\usepackage{tikz}
\newtheorem{lemma}{Lemma}

\newtheorem{prop}{Proposition}
\theoremstyle{definition}
\newtheorem{defn}{Definition}

\newtheorem{rem}{Remark}

\newtheorem{ass}{Assumption}
\begin{document}

\title{Epidemics with Behavior
%\thanks{We would like to thank}
}
\author{Satoshi Fukuda\thanks{Bocconi University, Department of Decision Sciences and IGIER. Email: \texttt{satoshi.fukuda@unibocconi.it}.} \and Nenad Kos\thanks{Bocconi University, Department of Economics, IGIER and CEPR. Email: \texttt{nenad.kos@unibocconi.it}.} \and Christoph Wolf\thanks{Bocconi University, Department of Economics and IGIER. Email: \texttt{christoph.wolf@unibocconi.it}.}}
\date{\today}

\maketitle
\thispagestyle{empty}

\begin{abstract}
We study equilibrium distancing during epidemics. Distancing reduces the individual's probability of getting infected but comes at a cost. It creates a single-peaked epidemic, flattens the curve and decreases the size of the epidemic. We examine more closely the effects of distancing on the outset, the peak and the final size of the epidemic. First, we define a behavioral basic reproduction number and show that it is concave in the transmission rate. The infection, therefore, spreads only if the transmission rate is in the intermediate region. Second, the peak of the epidemic is non-monotonic in the transmission rate. A reduction in the transmission rate can lead to an increase of the peak. On the other hand, a decrease in the cost of distancing always flattens the curve. Third, both an increase in the infection rate as well as an increase in the cost of distancing increase the size of the epidemic. Our results have important implications on the modeling of interventions. Imposing restrictions on the infection rate has qualitatively different effects on the trajectory of the epidemics than imposing assumptions on the cost of distancing. The interventions that affect interactions rather than the transmission rate should, therefore, be modeled as changes in the cost of distancing.\newline 

\end{abstract}

\clearpage
\pagenumbering{arabic}
\setcounter{page}{1}

\section{Introduction}
The earliest account of an epidemic is the Plague of Athens\textemdash an epidemic that tore through Athens 430 BC.\footnote{While termed the Plague, it is not clear what the disease was. Examination of DNA in dental pulp suggests it could have been typhoid fever; see \citet{Papagrigorakis_et_al_06}.} The epidemic played an important role in the weakening of Athens and the consequent victory of the Spartans; a detailed account is provided in \emph{History of the Peleponnesian war} by Thucydides. Outbreaks of epidemics have been a mainstay ever since, as has been their depiction in literature.\footnote{The Plague plays an important role in \emph{Decameron} by Boccaccio, \emph{Romeo and Juliet} by Shakespeare, and \emph{The Plague} by Camus. A case of smallpox is vividly depicted in Dickens' \emph{Bleak House}. Besides the disease, these works provide insights into how individuals behave to avoid the disease and what preventive steps authorities undertake. For example, in \emph{Romeo and Juliet} friar John is detained when trying to deliver Juliet's letter from Verona to Mantua. The movements between the two cities are limited to stop the spread of the plague.} Epidemics, however, are not merely a remnant of distant history or creative artistic representation. Not a quarter way into the 21st century, the world has faced outbreaks of Severe Acute Respiratory Syndrome (SARS), avian flu, swine flu, Ebola, Middle East Respiratory Syndrome (MERS), and COVID-19.

The systematic study of epidemics and how they spread through the population gained traction with the SIR model of \citet{Ross_Hudson_1917} and \citet{Kermack_McKendrick_27}. A disease is introduced into a population of susceptible individuals by a small seed of infected\textemdash and infectious\textemdash persons. The disease spreads through meetings of individuals. The infected eventually recover. By studying the transmission and the recovery rate of the disease, the model provides predictions on whether a disease will spread, %and in the case it does, 
what peak prevalence it will attain and what proportion of population will be afflicted. Yet, when it comes to individuals' behavior the model implicitly assumes that individuals engage in as many interactions at the height of the epidemic as they do when the disease is barely present. The model of \citet{Capasso_Serio_78} generalizes the standard model to non-linear or time-dependent contact rates, which can be thought of as a reduced form of modeling behavior; for a more recent overview, see \citet{Funk_et_al_10} and \citet{Verelst_et_al_16}. A strong point for explicitly modeling behavior is made by \citet{Ferguson_07}.

We propose a tractable model of equilibrium distancing during an epidemic.\footnote{Ours is not the first model of behavior during an epidemic. An account of the related literature follows below.} Susceptible individuals non-cooperatively decide whether, and to which extent, to distance or to engage in exposure at each point in time. Distancing comes at a quadratic cost, but reduces the probability of getting infected. The cost of getting infected is fixed; a model with fixed cost of infection was introduced by \citet{Chen_12}. We show that optimal distancing is proportional to the current size of the infected population. More infected individuals imply a higher likelihood that one gets infected and, therefore, more distancing. The difficulty in fully characterizing the equilibrium is that the prevalence of the disease depends on previous distancing behavior by all individuals. Nevertheless, we show that an equilibrium exists and that it is unique. Depending on parameters, two types of dynamics can arise. The disease either never takes off or it spreads through the population. In the latter case, the epidemic has a single peak: it propagates through the population until it reaches the peak prevalence, then it recedes and eventually dies out. Importantly, susceptible individuals distance throughout the epidemic, though the intensity of their distancing varies with the number of actively infected individuals. While distancing affects the entire path of the epidemic, it has important consequences for three crucial and commonly discussed measures: the conditions for an epidemic to start, its peak, and its final size. 

First, we define a basic reproduction number taking distancing into account\textemdash the \emph{behavioral basic reproduction number}. The behavioral basic reproduction number consists of the classical, epidemiological basic reproduction number, $R_0$, multiplied by a behavioral term.\footnote{We also derive an analogous \emph{behavioral effective reproduction number}. That the basic reproduction number itself may be misleading to understand epidemic dynamics has been recognized before. For example, \citet{caley2008} find that the observed attack rate of the 1918-1919 influenza pandemic was substantially lower than the one expected based on the basic reproduction number and attribute this discrepancy to social distancing.} We show that the disease propagates itself if and only if the behavioral basic reproduction number is larger than one. Interestingly, the behavioral basic reproduction number is concave in the transmission rate. Therefore, the disease spreads only for intermediate values of the transmission rate. This novel finding stands in stark contrast with the predictions offered by the SIR model without distancing \citep[see, for example,][]{brauer2012mathematical} where a disease spreads if its transmission rate is sufficiently high. With distancing, a higher transmission rate is required for the disease to take off as the caution of individuals lowers the behavioral basic reproduction number. However, if the transmission rate is too high, individuals distance with such fervor that the disease never takes off.\footnote{The importance of behavior at the outset of an epidemic for the entire trajectory of the epidemic is discussed, for example, in \citet{brauer2019final}.}

Second, we derive results pertaining to peak prevalence of the disease. Peak prevalence is of crucial importance to understand whether a disease might cause the health system to reach its capacity. For example, the 1918 influenza pandemic hit an unprepared health system which soon became overwhelmed \citep[see][]{jester2018100,schoch2001hospital}. In March 2020\textemdash less than a month after the coronavirus erupted in Italy\textemdash, the healthcare system in Northern Italy was under such severe pressure that some pneumonia patients could not be treated.\footnote{See https://www.nytimes.com/2020/03/12/world/europe/12italy-coronavirus-health-care.html.} Hence, the goal of many officials and societies became to \emph{flatten the curve} to avoid the active number of infected individuals exceeding the health care system's capacity.\footnote{\citet{kontreport} provide a calibration assessment of the risk of health system capacities being exceeded in the winter of 2020/21 due to COVID-19 patients in different European countries under various levels of lockdown effectiveness. They highlight that capacity constraints may become a severe concern again.} To improve the understanding of the effects that behavior has on peak prevalence, we study analytically how changes in the disease's transmission rate and changes in the cost of distancing affect the peak prevalence of a disease.

An increase in the cost of distancing unequivocally leads to a reduction in distancing and therefore to a higher peak prevalence of the disease. However, peak prevalence is non-monotonic in the transmission rate. If the transmission rate is high enough for the disease to spread but not too high, an increase in the transmission rate leads to an increase in the peak prevalence. In contrast, when the transmission rate is sufficiently high, an increase in the transmission rate decreases the peak prevalence and causes a flattening of the curve.\footnote{This finding is in accord with the result that for too high transmission rates individuals distance so much that the disease does not spread in the first place.} A decrease in the transmission rate\textemdash which in the SIR model without behavior always reduces peak prevalence\textemdash may have undesirable consequences in the short run by raising peak prevalence. Intuitively, there are two forces at play when the transmission rate is lowered: (i) conditional on behavior, the spread of the disease is slower, (ii) conditional on a particular prevalence, individuals face a lower infection risk and engage in less protective distancing. We show that the latter affect may outweigh the former leading to a higher prevalence of the disease.\footnote{This intuition is reminiscent of \emph{risk compensation} introduced by \citet{peltzman1975effects}: an intervention that improves an individual's security can lead to more risky behavior. For a survey, see \citet{hedlund2000risky}.} 

It is important to emphasize that the same cannot occur with a decrease in the cost of distancing. To the best of our knowledge, the body of work that studies non-pharmaceutical interventions models these either as reductions in the transmission rate \citep[see, for example,][]{Kruse_Strack_20,Rachel_20_Analytical} or as directly choosing the social activity level of individuals \citep[see, for example,][]{acemoglu2020multi,alvarez2020simple,Farboodi_et_al_20}\textemdash which are equivalent approaches in the SIR dynamics without behavior. Our results suggest that modeling individual distancing choices explicitly requires a careful choice of modeling interventions as qualitative implications differ through the behavioral channel. On the one hand, those interventions affecting the rate at which the disease propagates conditional on meetings, e.g., mandatory mask mandates, should be modeled as a decrease in the transmission rate.\footnote{Note that this result does not necessarily imply that mandating mask wearing in public spaces will worsen the epidemic; it may flatten the curve as well. However, we want to highlight the \emph{possibility} of this perverse effect arising. Indeed, \citet{chernozhukov_2021_causal} show that mask mandates have reduced the number of COVID-19 cases and deaths in the US.} On the other hand, those interventions that directly affect the incentives to distance, e.g., restaurant, bar or museum closures, should be modeled as a decrease in the cost of distancing.  Awareness of these potential detrimental effects of lowering the transmission rate is particularly important given that analytical results are rare and the majority of work on optimal policy is computational. Getting parameters in the calibration wrong can have adverse effects on the consequences of policy advice.

Third, we find that the possible detrimental short-run effects of a decrease in the transmission rate disappear in the long run. The total number of infected individuals throughout the epidemic is monotonically increasing in both the cost of distancing and the transmission rate. Conversely, the limiting number of susceptible individuals is decreasing in these parameters, but smaller than the ratio of recovery and transmission rate. In the SIR model without distancing, the number of infected individuals starts decreasing once the number of susceptible individuals is sufficiently low, in particular, once it falls below $\gamma/\beta$. When the number of susceptible individuals is too small, the pool of infected individuals is being depleted due to the rate of recovery being greater than the inflow of newly infected individuals. The number of susceptibles converges to a number strictly larger than $0$ and smaller than $\gamma/\beta$; for a derivation, see \citet{brauer2012mathematical}. Our model predicts a larger final size (more susceptibles, i.e., less total infections) than the standard SIR model due to distancing. Indeed, our model converges to the SIR model without distancing when the cost of distancing grows and so does the final size of the epidemic. Notably, as long as the cost of distancing is large enough for the disease to spread, the final size is below $\gamma/\beta$\textemdash even with distancing.

With these findings, we highlight an important trade-off between short-run mitigation, i.e., flattening the curve to avoid an overburdened health system, and long-run size of epidemics when considering the transmission rate. This trade-off arises due to the varying degree to which behavior matters during an epidemic. At the peak, the infection risks are high and individuals' distancing decisions have a strong impact on the dynamics of the epidemic. When an epidemic fades out, however, behavior is of less importance as individual risks are low and the standard SIR mechanics dominate the behavioral effects. However, the trade-off disappears once policies are considered that directly affect distancing incentives of individuals and both short-run mitigation and long-run size of the epidemic are obtained with similar policies, i.e., lowering the cost of distancing.

Next we provide bounds for the relevant values of cost of getting infected by studying a model in which the cost of infection is endegenously derived at each point in time. Finally, we provide a connection between models with explicit distancing decisions of individuals and models that generalize the contact and transmission rates of the standard SIR model, as in \citet{Capasso_Serio_78}. 

\textbf{Related literature.}  We provide a brief account of the related literature. We apologize for any omissions that might arise due to the speed with which new work is produced. \citet{Sethi_1978} introduces a behavioral component in an SIS model, but analyzes only the planner's problem. For subsequent references on behavior in SIS models, see \citet{Toxvaerd_19}. 

\citet{Reluga_10} proposes an SIR model with behavior and provides mostly numerical results. \citet{Chen_12} introduces an SIR model with a constant cost of infection, similar to what we analyze, but a more general contact function. His focus is on conditions on the contact functions that deliver uniqueness of the Nash equilibrium in each period for a given prevalence of the disease. \citet{Fenichel_et_al_11} and \citet{Fenichel_13} study a model in which the cost of getting infected is endogenous. They derive necessary conditions for equilibria and perform numerical analysis. \citet{Rachel_20_Analytical} and \citet{Toxvaerd_20} analyze a model of behavior with a linear cost of distancing and an endogenous time-varying cost of getting infected.\footnote{\citet{Rachel_20_Second} builds on this work to study lockdown effectiveness and the possibility of a second wave occurring.} They derive the necessary conditions for an equilibrium and derive two different paths that satisfy the necessary conditions. \citet{Farboodi_et_al_20} study a similar model numerically. \citet{Dasaratha_20} analyzes a model similar to ours where the infected individuals do not necessarily know whether they are infected. The complexity of his model requires that he mostly focuses on local results. Among other things, he shows that a marginal decrease in transmission rate around the peak of epidemic can lead to an increase in prevalence. \citet{McAdams_20_Nash} proposes a model in which an individual's benefit of social activities depends on the actions of other individuals and shows that there is a unique equilibrium of social activity choices in each period. \citet{McAdams_21} provides an excellent account of the rapidly growing literature. 

The majority of the literature on behavior and policy over an epidemic focuses on numerical results and simulations. \citet{makris2020great} numerically study how the expectation of the arrival of a pharmaceutical innovation affects individuals' optimal distancing. \citet{Toxvaerd_Rowthorn_20b} compare the individual and planner's decisions to apply treatments and vaccinations as pharmaceutical interventions during an epidemic. \citet{giannitsarouwaning} provide numerical projections for the COVID-19 pandemic, based on a model with endogenous distancing. \citet{acemoglu2020multi} analyze optimal lockdowns as a direct reduction of individuals' activity levels and calibrate the model to the COVID-19 pandemic in the US.

\section{The Model}\label{themodel}

We study behavior in an otherwise standard SIR model. A continuum of individuals, indexed by $i$ and normalized to unity, are infinitely lived with time indexed by $t\in [0,\infty)$. Each individual can be in one of the three states: susceptible, infected (and infectious), or recovered. Susceptible individuals might get infected in which case they transition into the infected state. Infected individuals can recover, but cannot become susceptible again. Recovered individuals acquire permanent immunity.\footnote{The standard SIR model is suitable for  viral diseases which are transmitted directly from human to human in a given period of time. Such viral diseases include measles, chickenpox (varicella), mumps, rubella, smallpox, influenza, poliomyelitis, Ebola virus disease, SARS, MERS, and COVID-19.} We denote the share of the population that is susceptible at time $t$ by $S(t)$, infected by $I(t)$ and recovered by $R(t)$. 

At each moment in time, susceptible individual $i$ chooses how much activity to engage in, denoted by $\varepsilon_i(t) \in [0,1]$. The individuals enjoy the activity, but it exposes them to the danger of infection; hence, termed exposure. The converse, $d_{i}(t):=1- \varepsilon_{i}(t)$, is the measure of distancing an individual engages in. While susceptible, the individual incurs a flow playoff, $\pi_{S}$. Distancing is uncomfortable and comes at the cost $\frac{c}{2}(d_{i}(t))^{2}$. Getting infected comes at a cost $-\eta$, $\eta<0$.

Individuals meet through a pairwise-matching technology where each individual has an equal chance of meeting any other individual\textemdash regardless of which state they are in. The only matches of consequence\textemdash that is, the only matches with an infection risk\textemdash are the ones between a susceptible and an infected individual. The probability that a susceptible individual who chooses exposure level $\varepsilon_i(t)$ meets an infected individual at time $t$ is $\varepsilon_i(t) I(t)$. The rate at which the infection is transmitted, upon such a match, is $\beta>0$. The rate at which a susceptible individual $i$ who chooses exposure $\varepsilon_i(t)$ gets infected is, therefore, $ \beta\varepsilon_i(t) I(t)$.\footnote{We implicitly assume that infected individuals choose full exposure. Though strong, the assumption is not as stark as it might at first seem. It is straightforward to accommodate exposure of infected with some parameter $e$, as long as it is fixed over time. Then, the same model as ours can be obtained by defining $\tilde \beta = e \beta$.} Finally, infected individuals recover at rate $\gamma>0$.\footnote{\label{fn:mortality}Following \citet{Keeling_Rohani_2008}, it is straightforward to incorporate fatalities from the disease into the SIR model by introducing a probability of death before recovery, $\sigma$, and rescaling the recovery rate $\gamma$ to $\gamma/(1-\sigma)$.} 

At each point in time $t$, a susceptible individual $i$ solves the problem
\begin{equation}\label{eq:naive_prob}
\max_{\varepsilon_{i}(t) \in [0,1]} \pi_{S} - \frac{c}{2} (1-\varepsilon_{i}(t))^{2}  +  \beta I(t) \varepsilon_{i}(t) \eta.
\end{equation}

Let $\varepsilon(t) := \frac{1}{S(t)}\int_{i \in S(t)}\varepsilon_i(t)di$ be the average exposure at time $t$. Analogously, define $d(t):=1-\varepsilon(t)$ as the average distancing at time $t$. Then, the model is governed by the following dynamics
\begin{align} 
\dot{S}(t)&=-\beta \varepsilon(t) I(t) S(t), \label{eq:S_dot}\\
\dot{I}(t)&= \beta \varepsilon(t)S(t)I(t)-\gamma I(t),  \label{eq:I_dot}\\
\dot{R}(t)&= \gamma I(t), \label{eq:R_dot}
\end{align}
with the assumption that there is a seed of infected, $I(0)=I_0\in(0,1)$, and susceptible individuals, $S(0)=S_{0}=1-I_{0}$. The number of susceptible individuals is ever shrinking. The number of infected individuals, on the other hand, is growing as long as the inflow of infections, $ \beta \varepsilon(t)S(t)I(t)$, is larger than the outflow of recoveries,  $\gamma I(t)$, and shrinking otherwise. It should be noted that $\dot{S}(t) + \dot{I}(t) = - \gamma I(t).$ The net inflow of individuals into the pool of susceptibles and infected is negative due to recoveries. Needless to say, since $S$, $I$ and $R$ are the only three states
\begin{align*}
S(t) + I(t) + R(t) = 1
\end{align*}
at each instance of time.

\begin{defn}
A symmetric equilibrium (an equilibrium, for short) is a tuple of functions $( S, I, R, (\varepsilon_{i})_{i \in [0,1]})$ with the following three properties: (i) $(S, I, R)$ follow (\ref{eq:S_dot}), (\ref{eq:I_dot}) and (\ref{eq:R_dot}) with the initial condition $(S(0),I(0),R(0)) = (S_{0},I_{0},0)$, where $\varepsilon$ is the average exposure; (ii) each $\varepsilon_{i}$ solves (\ref{eq:naive_prob}), that is, $\varepsilon_{i}$ is a best-response to $(S,I,R)$, where the average exposure $\varepsilon$ is induced by $(\varepsilon_{j})_{j \neq i}$; and (iii) $\varepsilon = \varepsilon_{i}$ for all $i \in [0,1]$.
\end{defn}

The first-order condition to the individual's problem yields the individual's optimal distancing choice
\begin{align}\label{eq:naive_distancing}
d_{i}(t) := 1- \varepsilon_{i}(t) = \min \left( - \frac{\eta \beta}{c} I(t), 1 \right).
\end{align}

When $- \frac{\eta \beta}{c} I(t)$ exceeds unity, individuals fully distance. Distancing at time $t$ depends only on the infected population at time $t$\textemdash up to constants $\beta$, $c$ and $\eta$. In a symmetric equilibrium, $\varepsilon_i = \varepsilon$ for all $i$. By equation (\ref{eq:naive_distancing}), exposure in a symmetric equilibrium is
\begin{equation*}%\label{eq:epsilon_naive}
\varepsilon(t) = \max \left( 1 + \frac{\eta \beta}{c} I(t),0 \right).
\end{equation*}
Plugging this equation into the SIR dynamics yields
\begin{align}
\dot{S}(t) & = - \beta S(t) I(t) \max \left( 1 + \frac{\eta \beta I(t)}{c},0 \right), \label{eq:naive_dot_S} \\
\dot{I}(t) & = \beta S(t) I(t) \max \left( 1 + \frac{\eta \beta I(t)}{c},0 \right) - \gamma I(t), \label{eq:naive_dot_I} \\
\dot{R}(t) & = \gamma I(t), \label{eq:naive_dot_R}
\end{align}
with the initial condition $(S(0),I(0),R(0)) = (S_{0}, I_{0}, 0)$ and $I_{0} = 1-S_{0} \in (0,1)$. Since this is an initial value problem, the following result obtains. All the proofs are collected in the Appendix.

\begin{prop}\label{prop:uniqueness_naive}
A symmetric equilibrium exists and is unique.
\end{prop}

\section{Analysis of the SIR Model with Behavior}

While the above system of differential equations does not have a tractable closed-form solution, we nevertheless establish qualitative properties of the equilibrium and the resulting epidemic dynamics. In this section, we proceed in four steps. First, we derive conditions for an outbreak to occur and show that, in that case, any epidemic is  single-peaked. Second, we derive an implicit equation for the solution path in the $(S,I)$-phase space. Third, we show how peak prevalence is affected by changes in the transmission rate, $\beta$, and the cost of distancing, $c$. Fourth, we derive how the final size of the epidemic is affected by the transmission rate and the cost of distancing.

\subsection{Single-peaked Epidemic and Its Onset}\label{sec:single_peak_onset}
We start by showing that the number of active cases peaks at most once.

\begin{prop}\label{prop:single_peak}
If $\hat t$ is such that $\dot I(\hat t) = 0$, then $\ddot I (\hat t) < 0$ and at $\hat t$ distancing is maximized.
\end{prop}

Proposition \ref{prop:single_peak} implies that if $I$ has a critical point, this critical point has to be a local maximum. Together with the continuous differentiability of $I$, this implies that $I$ can have at most one peak. The infection either immediately dies out, or becomes an epidemic with a single peak. Moreover, distancing is maximized at the peak of the epidemic.

In the standard SIR model, the infection propagates itself only if the basic reproduction number, $R_0:= \frac{\beta}{\gamma} S_{0}$, is larger than 1; see \citet{Heesterbeek_Dietz_96}.\footnote{Depending on the source $R_0$ is defined either as $\beta/\gamma$ or $\beta S_0/\gamma$. We use the latter definition as it allows for an easier presentation of results.} The basic reproduction number is the number of infections an infected person would cause during her infection if introduced into a population of susceptible individuals of size $S_0$. Namely, each infected causes $\beta S_0$ transmissions per unit of time while her infection lasts $1/\gamma$ periods on average. From a practical point of view, the observed and measurable variable is how many secondary infections have been caused given an individual's behavior. Therefore, incorporating the behavioral component in the basic reproduction number is of paramount importance. For a discussion of this issue\textemdash diseases that after an outbreak grows more slowly than expected\textemdash, see, for example, \citet{brauer2019final}. We define the \emph{behavioral basic reproduction number} as: 
\begin{align}\label{eq:r0d}
R_0^b := \frac{\beta }{\gamma}S_0\varepsilon(0).
\end{align}
Notice that $R_0^b =\varepsilon(0) R_0$. That is, the behavioral basic reproduction number consists of the virus-inherent basic reproduction number $R_0=\frac{\beta}{\gamma} S_{0}$ and a behavioral factor $\varepsilon(0)$. Equation (\ref{eq:I_dot}) at $t=0$ can now be rewritten as $\dot I(0) = \frac{I_{0}}{\gamma}(R_0^b-1)$. Therefore, the infection spreads, $\dot I(0) >0$, if and only if $R_0^b>1$, paralleling a similar resort in the model without distancing. However, while in the standard SIR model $R_0$ is increasing in $\beta$, the behavioral basic reproduction number $R_0^b$ is non-monotonic here and, in particular, concave. This follows from $\varepsilon(0) = 1+ \frac{\eta \beta}{c}I_0$ and (\ref{eq:r0d}); recall that $\eta <0$. This finding has important implications on which types of an infection will spread.

\bigskip
\begin{prop}\label{prop:I_0}
Fix $I_{0} \in (0,1)$. Then, $\dot I(0) > 0$ if and only if $R_0^b >1$. Moreover:
\begin{enumerate}
	\item \label{itm:hurlyburly1} If $I_{0} < \frac{1}{1-\frac{4 \eta \gamma}{c}}$, there exist $\underline \beta$ and $\overline \beta$, with $\frac{\gamma}{1-I_{0}} < \underline{\beta} < \overline{\beta} < - \frac{c}{\eta I_{0}}$, such that  $\dot{I}(0) >0$ if and only if $\beta \in (\underline{\beta}, \overline{\beta})$. In that case $\varepsilon$ admits an interior solution for all $t$. 
	\item \label{itm:hurlyburly2} If $I_{0} \geq \frac{1}{1-\frac{4 \eta \gamma}{c}}$, then $\dot{I}(t) \leq 0$ for all $t$. 
\end{enumerate}
\end{prop}

\begin{figure}
\begin{minipage}{0.49 \hsize}
\begin{center}
\includegraphics[width=\textwidth]{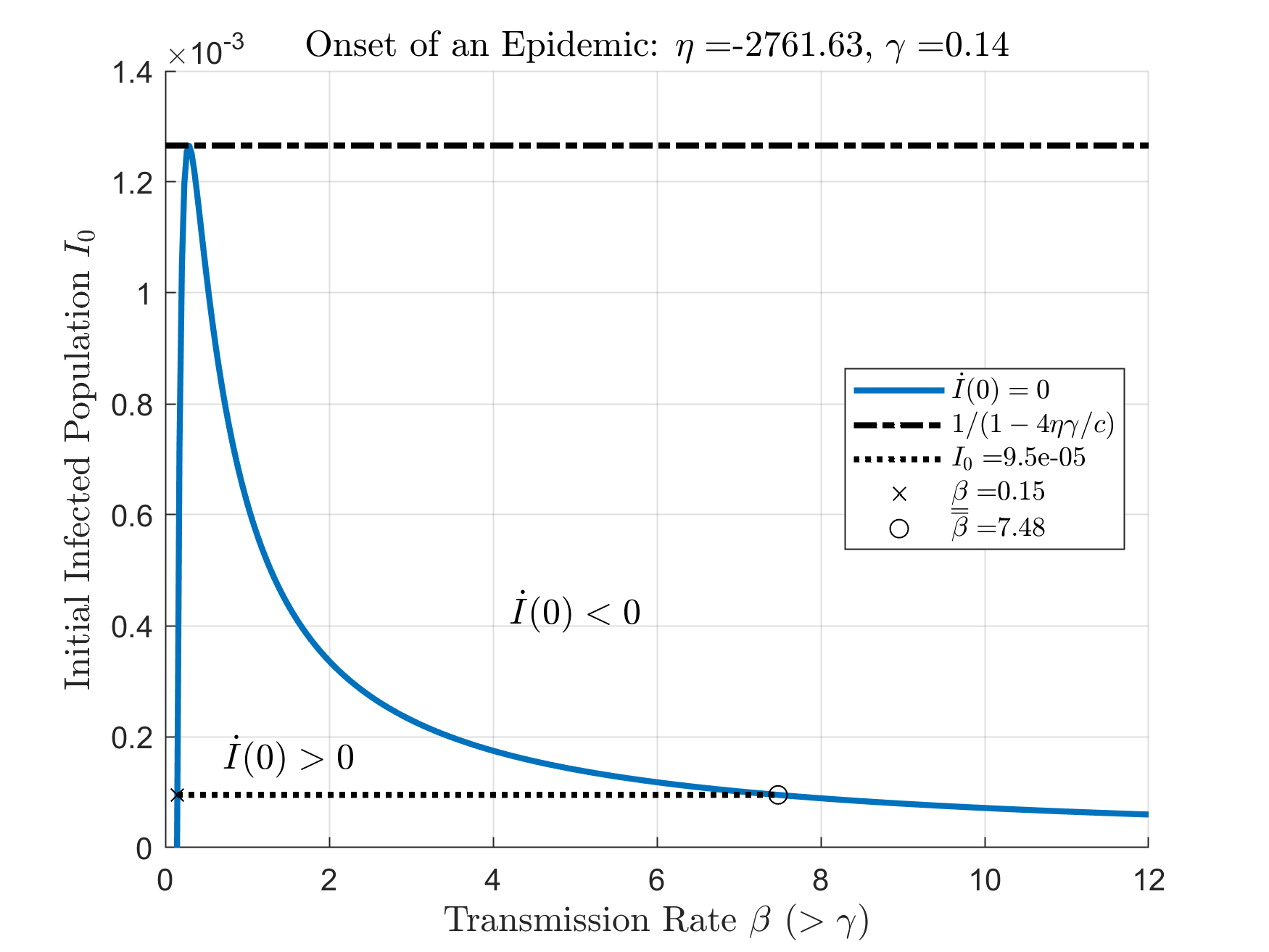}
\end{center}
\end{minipage}
\begin{minipage}{0.49 \hsize}
\begin{center}
\includegraphics[width=\textwidth]{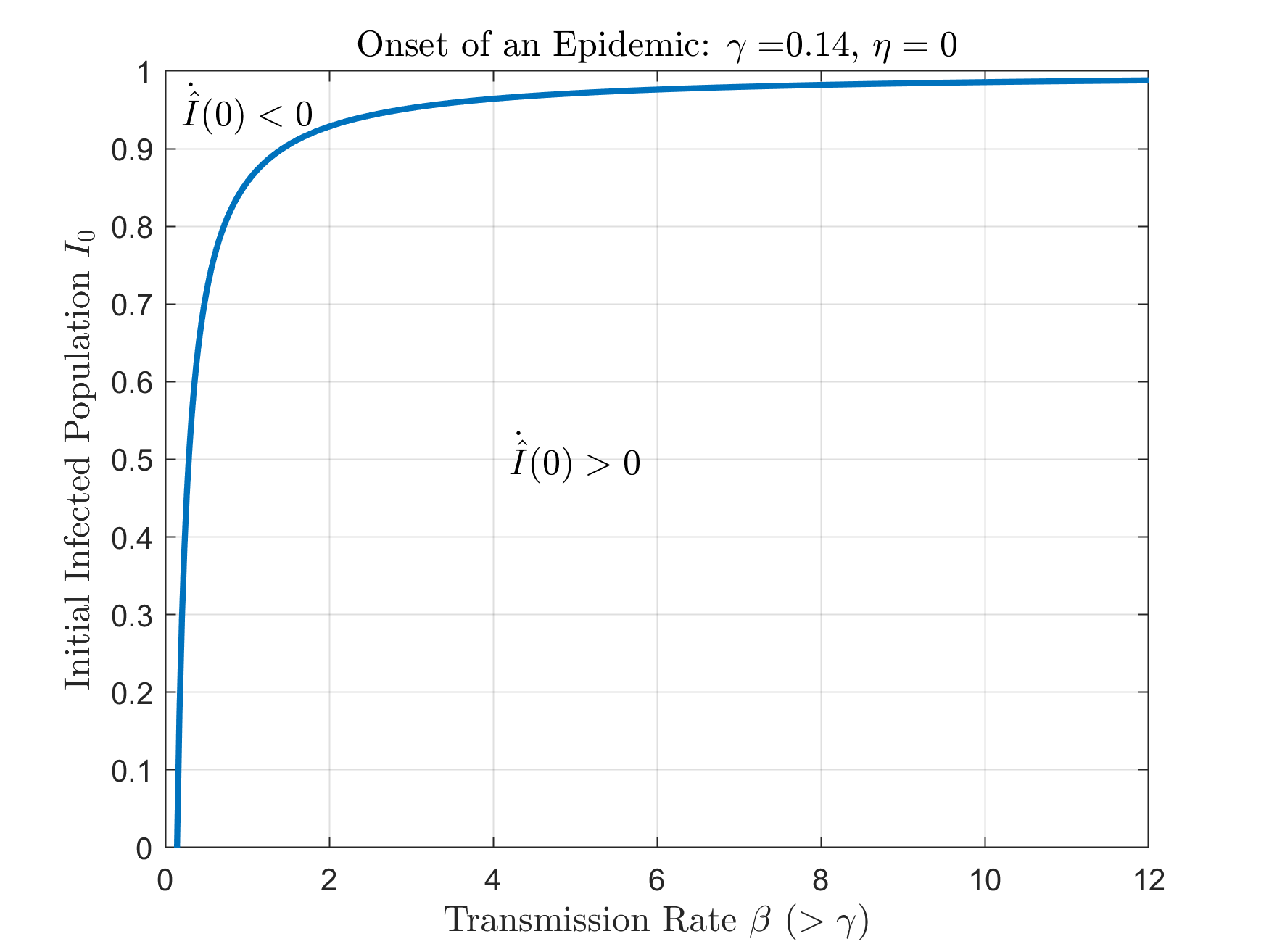}
\end{center}
\end{minipage}
\caption{\emph{The Onset of an Epidemic. Left: Constant cost of infection model; Right: Standard SIR model.} The solid line in each panel depicts the combination of $(\beta, I_{0})$ with $\dot{I}(0)=0$.} \label{fig:I_dot_0_naive}
\end{figure}

In the standard SIR model, for the infection to spread, $\beta$ must be high enough. In particular, $\beta > \frac{\gamma}{S_0}$. In the model with distancing, instead, the transmission rate has to be large enough to also overcome the initial distancing: 
\begin{equation*}
\beta >  \frac{\gamma}{(1-d(0)) S_0} > \frac{\gamma}{S_0}.\footnote{It should be noted that $d(0)$ depends on $\beta$ as well.}
\end{equation*}
Thus, our model predicts that a higher transmission rate is needed for the epidemic to start than in the standard SIR model. However, what differentiates the model with distancing from the standard SIR model even more starkly is that the infection peters out if the transmission rate is too high. If the disease is highly contagious, individuals are much more cautious, up to the point that their resolute distancing alone is sufficient to stop the disease in its tracks from the onset.\footnote{An informal discussion of the role of disease-intrinsic parameters and its effect on the outbreak of an epidemic can be found in \citet{christakis2020}.} In the face of preventive behavior the infection, therefore, spreads only if its transmission rate is large enough, but not too large, as illustrated in the left panel of Figure \ref{fig:I_dot_0_naive}.\footnote{We use parameters for COVID-19 in our simulations. A summary and justification of the parameters chosen can be found in Appendix \ref{app:parameters}. We also describe our numerical algorithm there.}

\begin{figure}[h!]
\begin{center}
\includegraphics[height=0.4\textheight]{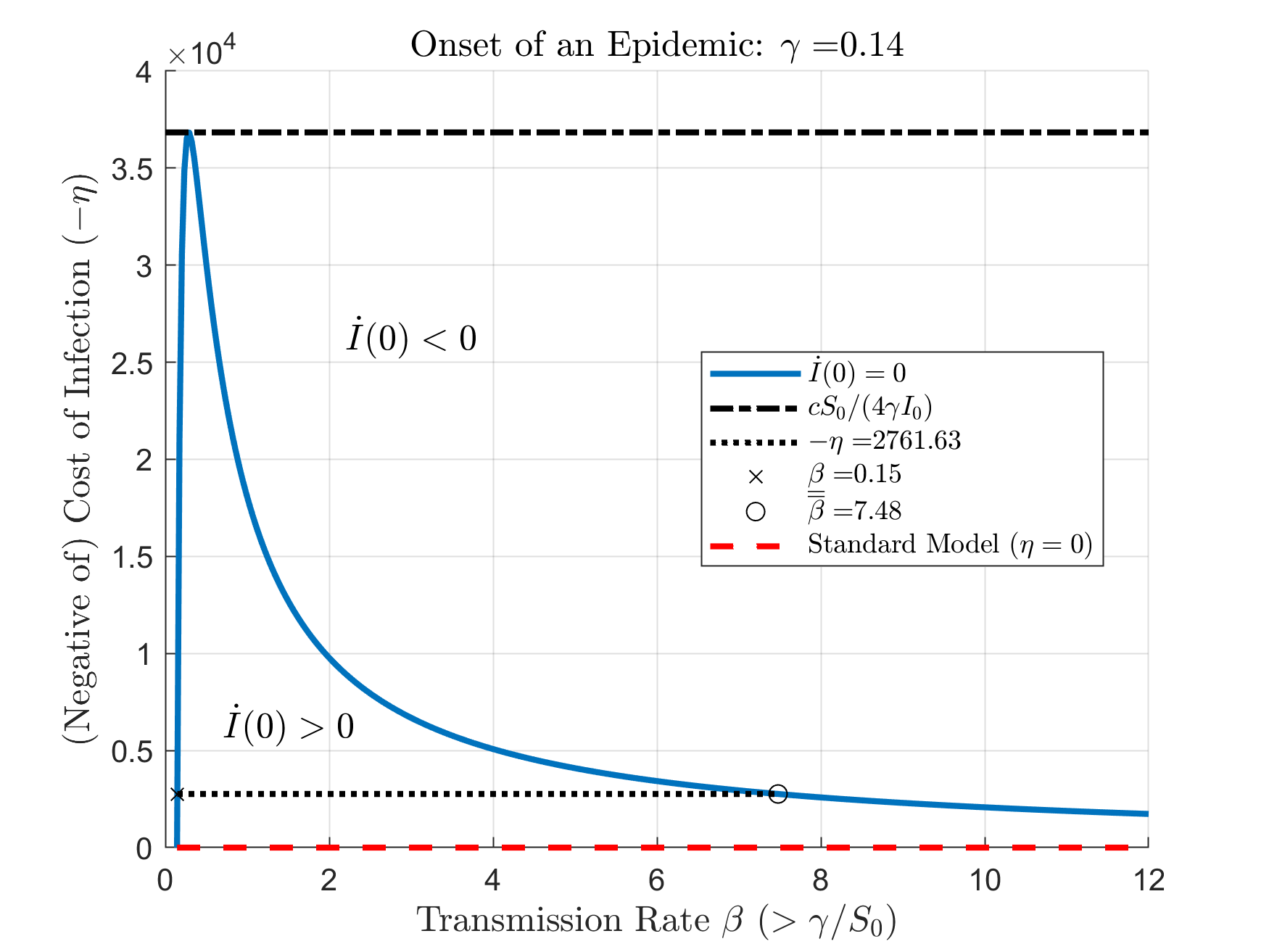}
\end{center}
\caption{\emph{The Onset of an Epidemic Along the Severity and Transmissibility Dimensions.} The solid line depicts the combination of $(\beta,-\eta)$ with $\dot{I}(0)=0$. The red dashed line depicts the set of transmission rates for the standard SIR model without behavior (which has $-\eta=0$) such that $\dot{I}(0)>0$.} \label{fig:I_dot_0_severity_transmissibility}
\end{figure}

Proposition \ref{prop:I_0} derives conditions on the transmission rate such that an epidemic takes off. The same question can be analyzed along other dimensions. For example, the CDC has adopted a categorization for influenza viruses along the severity-transmissibility dimensions \citep[see][]{reed2013novel}. In our model, this can be interpreted as categorizing the combination of the cost of infection, $-\eta$, and the transmission rate, $\beta$. The range of parameters $(\beta,-\eta)$ such that the disease takes off is depicted in Figure \ref{fig:I_dot_0_severity_transmissibility}.

Recall that in the SIR model without behavior\textemdash which is nested in our model as the case $\eta=0$\textemdash the epidemic takes off whenever $\beta S_{0}>\gamma$. This corresponds to the red dashed line in the figure. As the cost of the infection, $-\eta$, increases, individuals' distancing incentives start to matter for the onset of an epidemic. In particular, for a fixed $\beta$, the higher the cost of infection, the more individuals engage in distancing. If the cost of infection becomes very large, it prevents the disease from spreading altogether: $\dot{I}(0)<0$. There is a cutoff cost of infection such that the disease will never spread when $-\eta>\frac{c}{4\gamma}\frac{S_0}{I_0}$ as getting infected is so costly for individuals that their distancing behavior will compensate any transmission rate $\beta$.

It follows that a disease can only spread if its $(\beta,-\eta)$-combination is intermediate. For a given $\beta$, the infection cost must not be too high; while for a given $-\eta$, the transmission rate must neither be too high nor too low. The existence of an upper and a lower bound for $\beta$ follows the same intuition as the one applying for Proposition \ref{prop:I_0}.

\subsection{The Dynamics}

It is often useful to think of the model in terms of a phase diagram in the $(S,I)$-space, a graph showing how the number of infected individuals changes with the number of susceptible individuals. To find the solution path $(S,I):=(S(t),I(t))_{t \geq 0}$ in the phase space, one derives the quotient differential equation
\begin{align}\label{eq:phase_diff_eq_naive}
\frac{dI}{dS} = -1 + \frac{\gamma}{\beta}\frac{1}{S}\frac{1}{\max \left( 1+\frac{\beta \eta}{c}I, 0\right)}
\end{align}
by dividing equation (\ref{eq:I_dot}) with equation (\ref{eq:S_dot}) and using (\ref{eq:naive_distancing}) for $\varepsilon$. 

\begin{prop}\label{prop:phase}
Suppose $d(0) <1$.\footnote{The assumption is made for ease of exposition directly on $d(0)$; Formula (\ref{eq:naive_distancing}) provides the corresponding assumptions on primitives. Also, if $d(0)=1$, then individuals engage in full distancing up to some point, after which an equation analogous to (\ref{eq:phase}) determines the dynamics of the epidemic.}
%It means that exposure at time $0$ is an interior solution. As in the discussion of Proposition \ref{prop:I_0}, this holds when $\beta < - \frac{c}{\eta S_{0}}$. If $\varepsilon(0)=0$ as $\beta \geq - \frac{c}{\eta S_{0}}$, then $(S,I)$ vertically falls (i.e., $\dot{S}=0$ and $\dot{I}<0$) in the phase space until $I$ hits $I(t) = - \frac{c}{\eta \beta}$ at time $t \geq 0$. After $t$, $(S,I)$ follows (\ref{eq:phase}) (as discussed, $\dot{S}<0$ and $\dot{I}<0$).} 
The solution path $(S,I)$ is implicitly determined by 
\begin{align}\label{eq:phase}
S = \frac{\exp \left( \frac{\beta^{2}\eta}{2\gamma c} \left( S+I + \frac{c}{\beta \eta} \right)^{2} \right)}{\displaystyle \exp \left( \frac{\beta^{2}\eta}{2\gamma c} \left( 1 + \frac{c}{\beta \eta} \right)^{2} \right) \frac{1}{S_{0}} + 2\beta \sqrt{\frac{(-\eta)}{2\gamma c}} \int_{\beta \sqrt{\frac{-\eta}{2\gamma c}} \left( S+I + \frac{c}{\beta \eta}\right)}^{\beta \sqrt{\frac{-\eta}{2\gamma c}} \left( 1 + \frac{c}{\beta \eta}\right)} e^{-v^2} dv} .
\end{align}
\end{prop}

\begin{figure}[h!]
	\begin{center}
		\includegraphics[height=0.4\textheight]{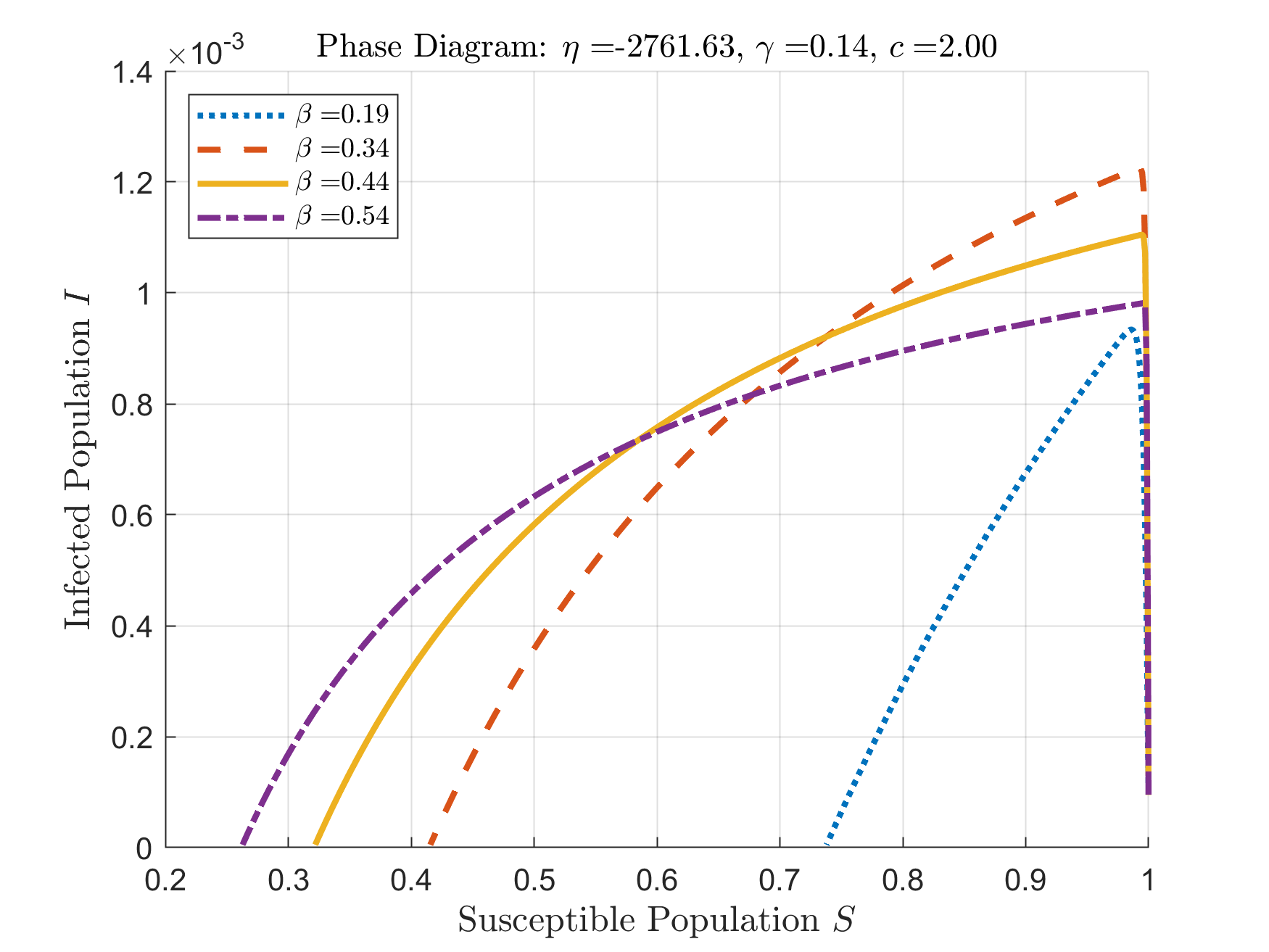}
	\end{center}
	\caption{\emph{Solutions Paths for Different Transmission Rates.}\label{fig:many_beta}}
\end{figure}

Figure \ref{fig:many_beta} depicts the solution paths in the phase space for different transmission rates. It is apparent that these are not monotonically ordered. 

To compare the dynamics of our model with the standard SIR model, denote by $(\hat{S}, \hat{I}, \hat{R})$ the proportion of susceptible, infected, and recovered individuals in the standard SIR model. We take the same initial condition $(\hat{S}(0), \hat{I}(0), \hat{R}(0)) = (S_{0}, I_{0},0)$ and the same parameters $(\beta, \gamma)$. The dynamics of the standard SIR model is obtained by replacing $(S,I,R,\varepsilon)$ with $(\hat{S}, \hat{I}, \hat{R}, 1)$ in equations (\ref{eq:S_dot}), (\ref{eq:I_dot}) and (\ref{eq:R_dot}). The solution path $(\hat{S},\hat{I})$ of the standard SIR model is captured by 
\begin{equation}\label{eq:phase_diff_eq_standard}
\frac{d\hat{I}}{d\hat{S}} = -1 + \frac{\gamma}{\beta}\frac{1}{\hat{S}}.
\end{equation}

\begin{figure}[h!]
	\begin{center}
		\includegraphics[height=0.4\textheight]{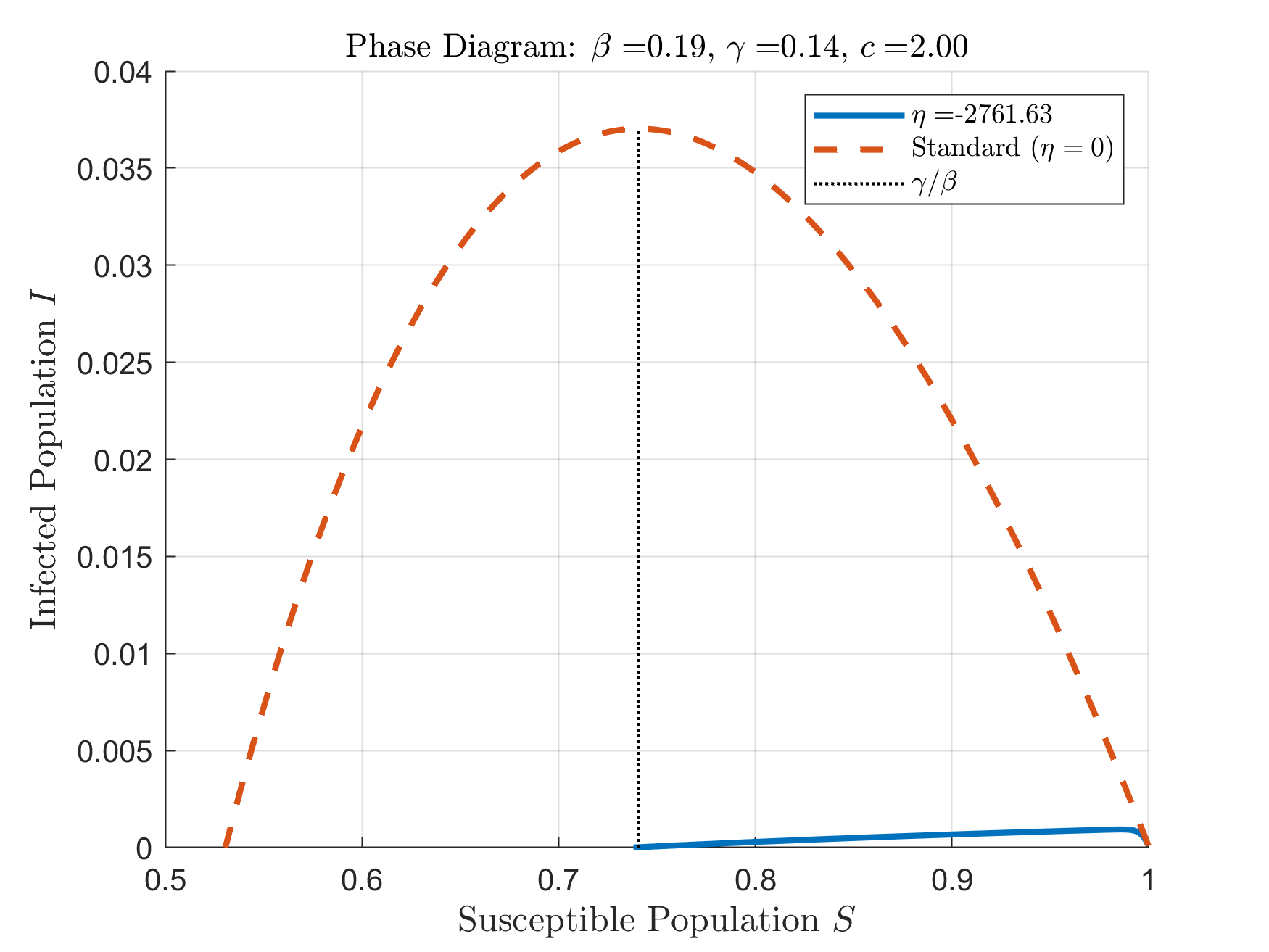}
	\end{center}
	\caption{\emph{Solution Paths for the SIR Models With and Without Behavior.} The solid curve depicts the solution path for our SIR model with behavior. The dashed curve depicts the solution path of the SIR model without behavior. The dotted vertical line depicts the herd-immunity threshold in the SIR model without behavior.} \label{fig:phase_compare}
\end{figure}

Comparing equations (\ref{eq:phase_diff_eq_naive}) and (\ref{eq:phase_diff_eq_standard}), we can show that the solution path $(\hat{S}, \hat{I})$ lies above the path $(S,I)$ in the phase space. More generally, we show that the solution path $(S,I)$ moves upwards as the cost of infection ($-\eta$) decreases. The SIR model without distancing can be recovered as the special case of our model with no cost of infection ($\eta = 0$). Hence, for any level of the susceptible population, the corresponding number of active infections is lower in the SIR model with distancing. These comparative statics are depicted in Figure \ref{fig:phase_compare}.

\bigskip
\begin{prop}\label{prop:phase_comparison}
Assume that $\dot{I}(0)>0$. The following hold:
\begin{enumerate}
\item \label{itm:prop_phase_comparison_1} As $\eta$ increases, the solution path $(S,I)$ moves upwards (lies above the original solution path). In particular, the solution path $(\hat{S}, \hat{I})$ of the standard SIR model, which is associated with $\eta=0$, lies above the original solution path.
\item \label{itm:prop_phase_comparison_2} Moreover, for every $t$, $S (t) \geq \hat{S}(t)$ and $R(t) \leq \hat{R}(t)$. 
\end{enumerate}
\end{prop}
 
Increasing the cost of infection, $-\eta$, pushes the solution path down, that is, it decreases the infected population at any level of susceptibles in the phase space. Since the SIR model without distancing can be seen as the special case of $\eta =0$, its solution path is above the path for any $\eta <0$. In particular, the peak prevalence in our model is below that of the standard model. 

While the first part of the above result compares the infected population $I$ for each level of the susceptibles $S$, the second part provides the comparison of the susceptibles $S$ and the recovered $R$ at each point in time. Namely, the susceptible population $S(t)$ in our model is always at least as large as in the standard SIR model while the recovered population $R(t)$ is at most as large as in the standard SIR model. The latter suggests that the cumulative infected population $\int_{0}^{t} I(s)ds = \frac{R(t)}{\gamma}$ in the model with behavior is smaller than in the standard SIR model. This result is intuitive as a higher cost of infection incentivizes individuals to take greater precaution in order to reduce their infection risk.

\subsection{Peak Prevalence}

The peak prevalence of an epidemic has profound consequences on the overall provision of health care services. A large number of infected individuals may lead to an overwhelming demand of personal protective equipment such as face masks and that of medical devices such as ICU beds and ventilators. The shortage of medical resources, in turn, may cause a suboptimal treatment and health care coverage.\footnote{As references regarding capacity constraints of the health care system see, for example, \citet{schoch2001hospital} for the 1918 influenza pandemic, \citet{ferguson2020report9} for the COVID-19 pandemic, and \citet{reed2013novel} for influenza epidemics.} The high demands of the epidemic on the health system also divert medical resources from other important activities such as vaccination against other infectious diseases and deliveries. What is more, health-care workers themselves are at high risk of infection.\footnote{\citet{Elston_et_al_17} survey the health impact of the 2014-15 Ebola outbreak in West Africa. For Siera Leone, they report a 20 \% decrease in measles coverage, an overall 20-23 \% decrease in deliveries and Caesarian sections. They also report that 10.7 \% of the health-care workforce were infected and 6.9 \% died from Ebola virus disease.} All this implies that the peak prevalence and how it is affected by parameters\textemdash in particular, by the disease's transmission rate but also by the individual's distancing cost\textemdash are of paramount interest for epidemic preparedness and optimal policy responses. 

When the epidemic takes off ($\dot{I}(0)>0$), the prevalence is maximized when $\dot I(t)=0$\textemdash Proposition \ref{prop:single_peak} implies that the local maximum is also global\textemdash that is, when 
\begin{align}
\varepsilon(t) S(t) = \gamma/\beta. \label{eq:peak_prevelance}
\end{align}
Let $I^{\ast}:= \max_{t}I(t)$.  % and $t^{\ast}$ be defined by $I(t^{\ast}) = I^{\ast}$.
Note that $I^\ast$ together with the corresponding $S^\ast$ is a solution to (\ref{eq:phase}) and (\ref{eq:peak_prevelance}). Despite this system of equations being intractable, the phase diagram analysis allows for several important comparative statics.

In the standard SIR model with $R_{0} >1$, the peak prevalence $\hat{I}^{\ast} := \max_{t} \hat{I}(t)$ is given by $\hat{I}^{\ast} = 1-\frac{\gamma}{\beta} + \frac{\gamma}{\beta} \log \left( \frac{\gamma}{\beta S_{0}} \right)$; see, for example, \citet{brauer2012mathematical} or \cite{Hethcote_2008}. The peak prevalence is attained when the population $\hat{S}(t)$ of susceptibles reaches the threshold of herd immunity $\frac{\gamma}{\beta}$. When the peak prevalence $I^{\ast}$ of our model is attained, the population $S(t)$ of susceptibles is larger than $\frac{\gamma}{\beta}$. Since the solution path $(S,I)$ is below the path $(\hat{S}, \hat{I})$, our model has a smaller peak prevalence than the SIR model without behavior, $I^{\ast} < \hat{I}^{\ast}$. 
We study how the peak prevalence changes with changes in the parameters $\beta$ and $c$. To focus on the case in which the infection can take place, we assume $I_{0} < \frac{1}{1-\frac{4\eta\gamma}{c}}$ so that $\dot{I}(0)>0$ if and only if $\beta \in (\underline{\beta}, \overline{\beta})$; see Proposition \ref{prop:I_0}. When this assumption fails, the infection dies out so that $I^{\ast}=I_{0}$.

\bigskip
\begin{prop}\label{prop:I_peak}
The following hold:
\begin{enumerate}
\item \label{itm:prop_I_peak_1} Fix $\gamma$, $c$ and $\eta$ and let $I_{0} < \frac{1}{1-\frac{4\eta\gamma}{c}}$% with $S_{0}=1-I_{0}$
. Then, $I^{\ast}$ is non-monotonic in $\beta \in (\underline{\beta}, \overline{\beta})$. In particular, there exist $\beta_1 < \beta_2$ such that $I^{\ast}$ is increasing in $\beta$ for $\beta \in (\underline \beta, \beta_1)$ and decreasing in $\beta$ for $\beta \in (\beta_2, \overline \beta)$.
\item \label{itm:prop_I_peak_2} The peak prevalence $I^{\ast}$ is non-decreasing in $c$. It is strictly increasing in $c$ whenever $\dot{I}(0)>0$.
\end{enumerate}
\end{prop}

\begin{figure}
\begin{minipage}{0.49 \hsize}
\begin{center}
\includegraphics[width=\textwidth]{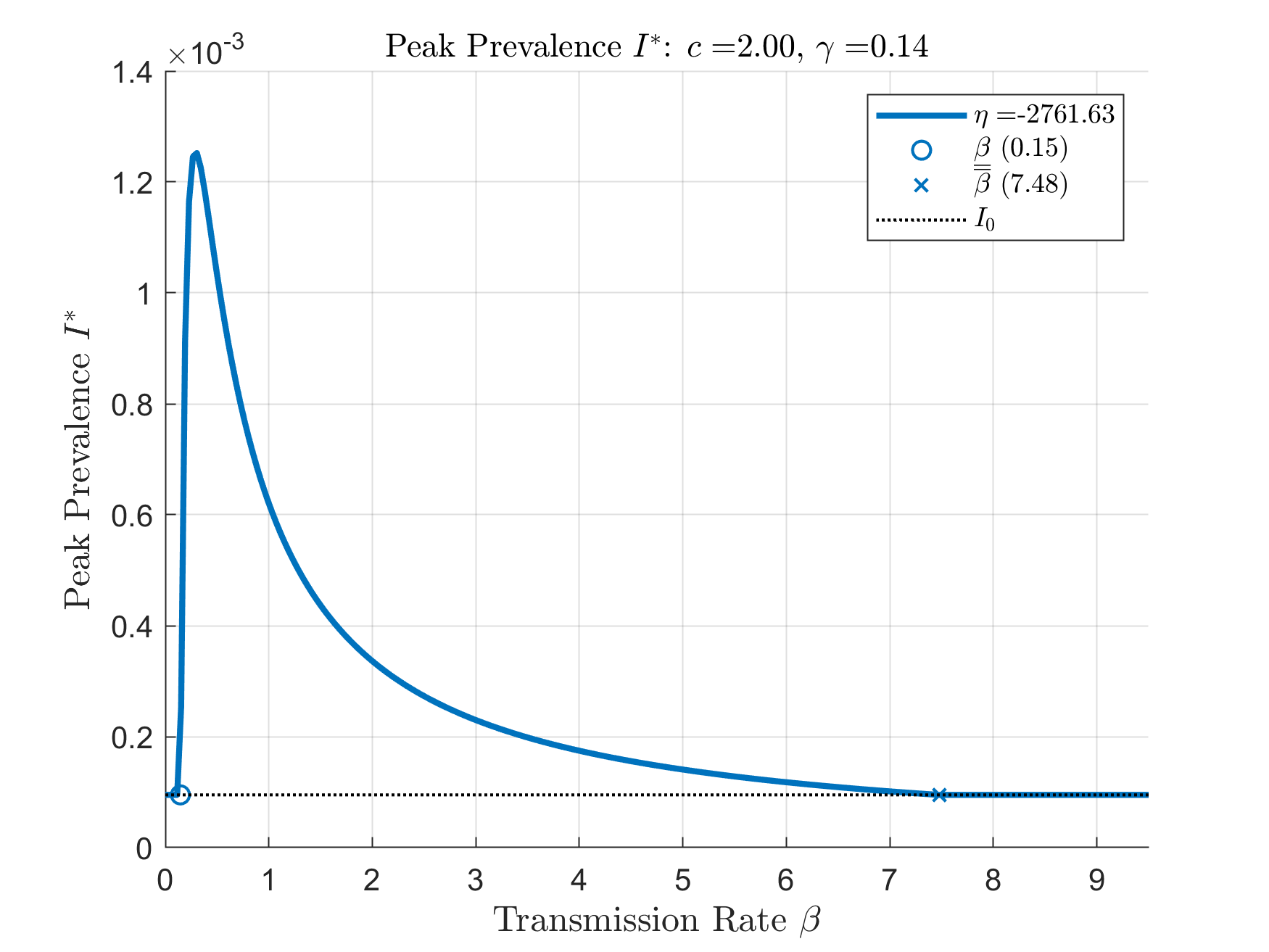}
\end{center}
\end{minipage}
\begin{minipage}{0.49 \hsize}
\begin{center}
\includegraphics[width=\textwidth]{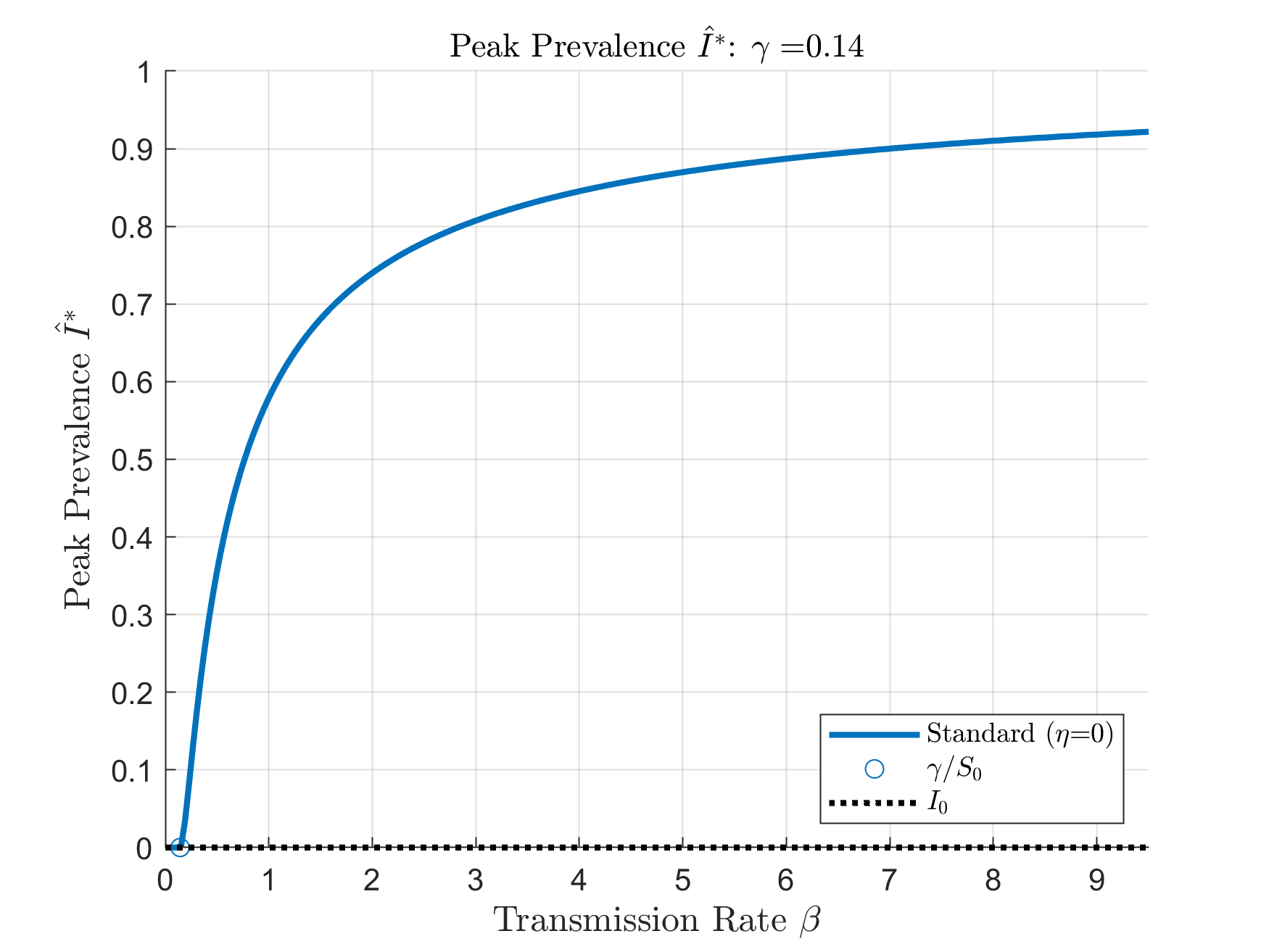}
\end{center}
\end{minipage}
\caption{\emph{Peak Prevalence as Function of Transmission Rate for the SIR Models with and without Behavior.} The left panel depicts peak prevalence of the SIR model with behavior. Initially, a higher transmission rate increases the peak. However, beyond a certain threshold, the peak decreases in the transmission rate as the distancing behavior outweighs the direct effect of a higher transmission rate. The right panel depicts peak prevalence of the SIR model without behavior.} \label{fig:peak_beta}
\end{figure}

In the SIR model without behavior, the peak prevalence $\hat{I}^{\ast}$ is monotonically increasing in the transmission rate $\beta$, as illustrated in the right panel of Figure \ref{fig:peak_beta}: the more transmissive a virus is, the higher the peak prevalence, provided that the infection takes off. In contrast, in our model, a higher rate of transmission leads to, ceteris paribus, more distancing. This effect can be so strong that a higher transmission rate reduces the peak prevalence and flattens the infection curve. More precisely, the above result establishes a non-monotonicity of peak prevalence in $\beta$. When the rate of transmission is low, the peak of the infection is increasing in $\beta$. However, when the rate of transmission is high, the peak prevalence decreases with $\beta$. The left panel of Figure \ref{fig:peak_beta} illustrates this non-monotonicity of the peak in the transmission rate. Proposition \ref{prop:I_peak} leads to a rather striking conclusion: a measure imposed to fight the epidemic through a decreased $\beta$ could have a daunting short-run effect; for example, if the potential resulting increase in prevalence leads to stress of the health care system. We want to emphasize that this effect arises only for a subset of potential parameters. In particular, \citet{chernozhukov_2021_causal} show that the introduction of mask mandates, which we consider to be a $\beta$-reducing policy, reduced the number of active cases in 2020 during the COVID-19 pandemic in the US. At the same time, \citet{knotek2020consumers} report survey evidence that some individuals view mask wearing as a substitute for physical distancing. This issue has been discussed in \citet{Howarde2014564118}.\footnote{This indirect effect of a measure reducing individual risk on taking less precautions is reminiscent of risk compensation. See, for example, \citet{peltzman1975effects} and \citet{hedlund2000risky}.}

On the other hand, an increase in the cost of distancing always increases the peak prevalence. Not surprisingly, a higher cost of distancing leads to less distancing, all else equal. The disparity in effects of changes in $c$ and $\beta$ on peak prevalence can be most readily seen by studying how the slope of the solution path at a fixed point in the phase space varies with changes in the two parameters. Differentiating the slope with respect to the cost of distancing parameter yields
\begin{align*}
\frac{\partial}{\partial c} \left( \frac{dI}{dS}\right) =-\frac{\gamma}{\beta \varepsilon^2 S} \frac{\partial \varepsilon}{\partial c}<0, 
\end{align*}
where the inequality follows from the observation that for a fixed $I$ the exposure increases if the cost of distancing increases. Importantly, the only effect an increase in the cost has on the solution path is through the change in distancing. By an implication, the slope of a solution path with a higher cost is smaller than the slope of a solution path with a smaller cost of distancing at any point of intersection. The fact that they start from the same point, $(S_0, I_0)$, then implies that everywhere else the solution path corresponding to a higher cost must be above the one with the lower cost. 

The change in the transmission rate, on the other hand, has a more nuanced effect. Differentiating the slope of the solution path at a fixed point yields
\begin{align*}
\frac{\partial}{\partial \beta}  \left( \frac{dI}{dS}\right) =-\frac{\gamma}{\beta^2 \varepsilon S} -\frac{\gamma}{\beta \varepsilon^2 S}  \frac{\partial \varepsilon}{\partial \beta},
\end{align*}
revealing that an increase in $\beta$ has two effects. A higher $\beta$ makes the infection more perilous. The increase in the transmission rate, holding everything else fixed, results in more secondary infections from each infected individual, thereby increasing the velocity of the spread of the disease. Such a direct effect is absent from changes in the cost of distancing. The second, indirect, effect is due to the response of distancing to the change in the transmission rate. A more infectious disease results in more distancing and thus dampens the evolution of the epidemic. The two effects run in opposite directions. Depending on which of the two dominates, an increase in $\beta$ can lead to either a smaller or a larger slope of the solution path. 

The above finding has an important implication on how various preventive policies should be studied in models with an epidemiological component. Such models commonly adopt two apparatus: behavior is either modeled implicitly by changes in $\beta$ in the standard SIR model or by directly imposing behavioral changes in models with behavior.\footnote{Examples of models of the former include \citet{brauer2019final,Capasso_Serio_78,Kruse_Strack_20} while examples of the latter include  \citet{acemoglu2020multi,alvarez2020simple,Farboodi_et_al_20,Rachel_20_Analytical}.} Our results point to the importance of differentiating between changes in the transmission rate and changes in the cost of distancing. For example, if a government imposes temporary restaurant closures to slow the spread of the disease, this gives individuals fewer reasons to go out and should be modeled as a decrease in the cost of distancing, and not as a decrease in the transmission rate.\footnote{Note that one can view holidays or vacation periods as an \emph{increase} in the cost of distancing which affects individual behavior as well.}

\subsection{Final Size of the Epidemic}

In the long run, the epidemic dies out, $\displaystyle I_{\infty} := \lim_{t \rightarrow \infty} I(t) =0$. After $S$ falls below $\gamma/\beta$, so does $\varepsilon S$, necessitating a reduction in the incidence rate. An important long-run property of the disease is $\displaystyle S_{\infty} := \lim_{t \rightarrow \infty} S(t)$, the number of remaining susceptible individuals once the epidemic is over. Note that $S_{\infty}$ is well-defined because $S$ is weakly decreasing. Conversely, $1-S_{\infty}$ is the size of the epidemic. In the SIR model without distancing, $\hat{S}_{\infty} := \displaystyle \lim_{t \rightarrow \infty} \hat{S}(t) \in \left( 0, \frac{\gamma}{\beta} \right)$.\footnote{In particular, see e.g., \citet{brauer2012mathematical} or \citet{Hethcote_2008}, $1-\hat{S}_{\infty} = \frac{\gamma}{\beta} \log \left( \frac{S_{0}}{\hat{S}_{\infty}} \right)$.} At the end of the epidemic, a strictly positive fraction of the population remains susceptible, $\hat{S}_{\infty} > 0$. The upper bound on this number is given by $\gamma/\beta$. This follows immediately in the standard model without distancing as the number of infected individuals is increasing whenever $\hat{S}(t) > \frac{\gamma}{\beta}$. Only after this threshold has been reached, the share of susceptibles falls and the epidemic starts to falter. The following result establishes how the model with distancing compares with respect to the size of the epidemic.

\begin{prop}\label{prop:epidemic size}
The following chain of inequalities holds:
\begin{align*}
0< S_{0} e^{- \frac{\beta}{\gamma}} \leq \hat{S}_{\infty} \leq S_{\infty} < \frac{\gamma}{\beta}.
\end{align*}
\end{prop}

\begin{figure}[h!]
	\begin{center}
		\includegraphics[height=0.4\textheight]{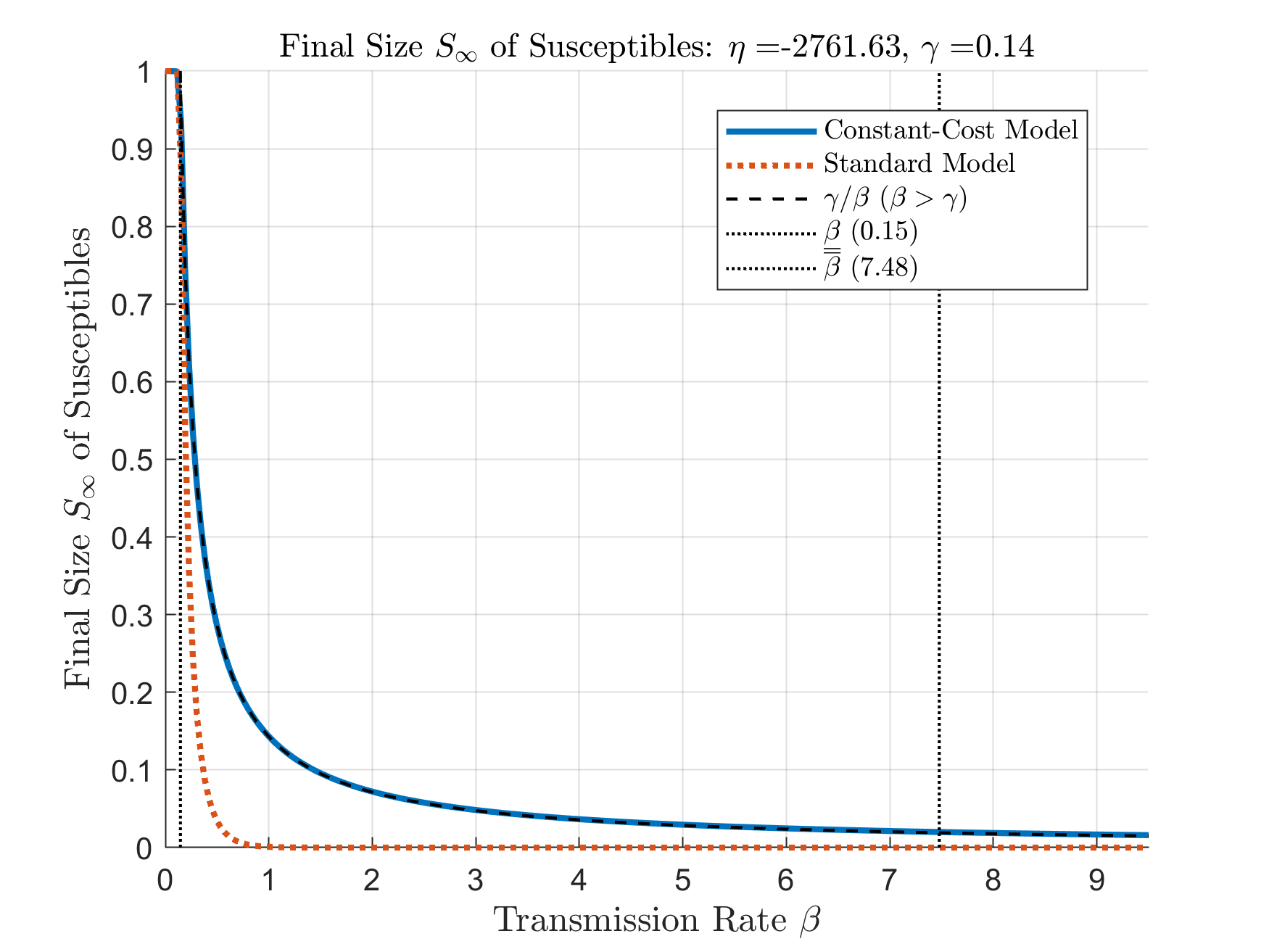}
	\end{center}
	\caption{\emph{Final Size of Susceptibles as Function of Transmission Rate.} The solid blue curve depicts the final size of susceptibles in our SIR model with behavior. The red dotted line shows as a comparison the final size in the SIR model without behavior which lies always below the solid blue curve. The dashed black line depicts the herd-immunity threshold that is the upper bound for the final size of susceptibles in both models.}\label{fig:final_size}
\end{figure}

Not all too surprisingly, the model with distancing predicts a smaller final size of the epidemic, $1-S_{\infty}$, than the model without distancing. More surprising is the fact that\textemdash even with distancing\textemdash the final number of susceptibles, $S_{\infty}$, cannot exceed $\gamma/\beta$. The reason is that in the limit $I$ tends to $0$. If $S_{\infty}$ was strictly above $\gamma/\beta$, then, as the epidemic would be winding down, so would the distancing. Since the exposure would be close to one, $\varepsilon S \approx S > \gamma/\beta$. But then (\ref{eq:I_dot}) implies that the epidemic should reignite, contradicting the supposition that it was winding down.

The following result studies the effect of the transmission rate and the cost of distancing on the final size of the epidemic and is graphically represented in Figure \ref{fig:final_size}.\\

\begin{prop}\label{prop:S_infty_monotone}
$S_{\infty}$ is decreasing in $\beta$, for $\beta \in [0, - \frac{c}{\eta I_{0}}]$, and $c$.
\end{prop}

The size of the epidemic, $1-S_{\infty}$, is monotone in $\beta$ and $c$, as long as $\dot I(0) > 0$.\footnote{To see this, note that $\overline{\beta}<- \frac{c}{\eta I_{0}}$.} Higher $\beta$ leads to a larger size of infection, as does an increase in the cost of distancing. The monotonicity of the final size of the epidemic in the transmission rate is in contrast with the result that the peak prevalence is non-monotonic in the same parameter. While policies that affect $\beta$ might have perverse effects in the short run\textemdash e.g., a decrease in the transmission rate, $\beta$, may lead to an increase in peak prevalence\textemdash the effects in the long run are desirable. If the hospital capacities are not at their limits, such policies will achieve the desired result, a reduction in cumulative infections, in the long run. In the short run, however, one needs to be circumspect if the medical capabilities are at or close to the capacity and a trade-off between short-run prevalence and long-run epidemic size might occur.

Intuitively, the effect of changes in the transmission rate on the final epidemic size is monotone and resembles the comparative statics of the standard SIR model because when the epidemic vanishes, i.e., when $I(t)$ approaches 0, individuals stop distancing. Hence, the only effect that changes in $\beta$ have in this final phase of the epidemic is the direct effect on infections.

\section{Endogenous Cost of Infection}\label{sec:endogensou_cost_infection}

In this section, we present a model with an endogenous cost of infection and establish a connection to the model with a constant cost of infection. As before, the individuals at each point in time decide how much to distance, which determines how likely they are to get infected. An individual's flow payoff from being in state $\theta \in \{S,I,R\}$ is $\pi_\theta$. We assume $\pi_{S} \geq \pi_{R} \geq \pi_{I}$.\footnote{Models with endogenous cost of infection have been presented in \citet{Reluga_10,Fenichel_et_al_11,Fenichel_13,McAdams_20_Nash,Rachel_20_Analytical,Toxvaerd_20}, among others. Yet, analytical characterizations of equilibria are rather elusive.} The endogeneity of costs of infection results from differences in the flow payoff across the states and the individual taking future infection risks into account. The individual discounts the future at rate $\rho >0$.

A susceptible individual $i$ with exposure $\varepsilon_{i}(t)$ enjoys the instantaneous payoff $\pi_S - \frac{c}{2}(1-\varepsilon_{i}(t))^2$. Let $p_i(t)$ be the probability of being infected at time $t$. Then,
\begin{align}\label{eq:p_dot}
\dot{p}_i(t) = \varepsilon_i(t) \beta I(t)(1-p_i(t)),
\end{align}
with $p_i(0) = 0$.\footnote{We model the behavior of susceptible individuals. Hence, the probability that they are infected at the outset is zero.} Once an individual gets infected, her progression to recovery is independent of her behavior.  Her continuation payoff from the moment she became infected, $V_I$, is:\footnote{Suppose that an individual gets infected at time $\tau$. The (conditional) probability that the individual will have been recovered after time $\tau + t$ is $1-e^{-\gamma t}$. Therefore, $$V_{I}(\tau) = \int_{0}^{\infty} e^{-\rho t} \left( e^{-\gamma t} \pi_{I} + (1-e^{-\gamma t}) \pi_{R} \right)dt,$$ which is independent of $\tau$. See also \citet{Toxvaerd_20}.}
\begin{align}\label{eq:Vi}
V_I = \frac{1}{\rho+ \gamma}\left( \pi_I + \frac{\gamma}{\rho} \pi_R\right).
\end{align}

A susceptible individual who faces average exposure $\varepsilon$ from her peers solves the problem
\begin{align}
& \max_{\varepsilon_{i}(\cdot) \in [0,1]} \int_{0}^{\infty} e^{-\rho t} \left\{ (1-p_{i}(t)) [\pi_{S}- \frac{c}{2}(1-\varepsilon_{i}(t))^{2}] + p_{i}(t)\rho V_{I}  \right\} dt \label{eq:ind_problem} \\
& \text{s.t.} \notag \\
& \dot{p}_{i}(t) = \beta \varepsilon_{i}(t) I(t) (1-p_{i}(t)), \notag \\
&  p_{i}(0)=0, \notag
%& \dot{S}(t) = - \beta \overline{\varepsilon}(t) S(t)I(t), \\
%& \dot{I}(t) = \beta \overline{\varepsilon}(t) S(t) I(t) - \gamma I(t) = I(t)(\beta \overline{\varepsilon}(t) S(t) - %\gamma), \\
%& \dot{R}(t) = \gamma I(t),
\end{align}
and the underlying dynamics given by (\ref{eq:S_dot}), (\ref{eq:I_dot}) and (\ref{eq:R_dot}) with the initial condition $(S(0),I(0),R(0))=(1-I_0,I_0,0)$ and $I_0\in(0,1)$.\footnote{The payoff in (\ref{eq:ind_problem}) can be obtained from 
\begin{align*}
\int_0^{\infty} e^{-\rho t}  (1-p_{i}(t)) \left[\pi_{S}- \frac{c}{2}(1-\varepsilon_{i}(t))^{2} + \frac{\dot p_i(t)}{1-p_i(t)} V_I \right] dt.
\end{align*}
With probability $1-p_i(t)$ individual $i$ has not been infected by time $t$ and receives the flow payoff $(\pi_{S}- \frac{c}{2}(1-\varepsilon_{i}(t))^{2})dt$. In addition, with probability $\dot p_i(t)dt$ she becomes infected and receives the lump sum payoff $V_I$. The above payoff is obtained by integration by parts. This approach was used in \citet{Toxvaerd_20}. For the approach dealing with all three states ($S$, $I$ and $R$) see \citet{Rachel_20_Analytical}.} The individual's payoff can be thought of as the expected value of being susceptible or infected at each point in time where the flow payoff of an infected individual is $\rho V_I$. An individual's behavior affects her probability of infection directly, but none of the population dynamics as she is small.

We study symmetric equilibria. 
\begin{defn}
A symmetric equilibrium (an equilibrium, for short) is a tuple of functions $( S, I, R, (\varepsilon_{i}, p_{i})_{i \in [0,1]})$ with the following three properties: 
\begin{enumerate}[(i)]
\item $(S, I, R)$ follow (\ref{eq:S_dot}), (\ref{eq:I_dot}) and (\ref{eq:R_dot}) with the initial condition $(S(0),I(0),R(0)) = (S_{0},I_{0},0)$, where $\varepsilon$ is the average exposure; \item each $\varepsilon_{i}$ solves (\ref{eq:ind_problem}), that is, $\varepsilon_{i}$ is a best-response to $(S,I,R)$, where the average exposure $\varepsilon$ is induced by $(\varepsilon_{j})_{j \neq i}$; and 
\item $\varepsilon = \varepsilon_{i}$ for all $i \in [0,1]$. 
\end{enumerate}
In equilibrium, each $p_{i}$ is determined by the average exposure $\varepsilon$ and $I$, and thus $p=p_{i}$ for each $i \in [0,1]$. For ease of exposition, we denote an equilibrium by $(S, I, R, \varepsilon, p)$.
\end{defn}

\bigskip
\begin{ass}
$\pi_S - \frac{c}{2} > \rho V_I$.
\end{ass}
Even if a susceptible individual is fully distancing, her flow payoff of being suceptible is greater than the flow payoff of being infected. Differently, the infection is so severe that an individual faced with the choice between fully distancing forever and being infected with certainty chooses the former.

The current-value Hamiltonian of problem (\ref{eq:ind_problem}) is
\begin{align}\label{eq:hamilton}
\mathcal{H}_{i} & = (1-p_{i}(t))[\pi_{S} - \frac{c}{2}(1-\varepsilon_{i}(t))^{2}] + p_{i}(t)\rho V_{I} + \eta_{i}(t)\beta \varepsilon_{i}(t) I(t)(1-p_{i}(t)),
\end{align}
where $\eta_i(t)$ is the current-value co-state variable. It represents the marginal value of an increase in the probability of being infected at time $t$. The optimality condition with respect to exposure $\varepsilon_{i}(t)$ at time $t$ is
\begin{equation*}
\frac{\partial \mathcal{H}_{i}}{\partial \varepsilon_{i}(t)} = (1-p_{i}(t)) [ c(t) (1-\varepsilon_{i}(t)) + \beta \eta_{i}(t) I(t) ] =0.
\end{equation*}
Assuming that $p_{i}(t) < 1$, which will be verified in Remark \ref{rem:pless1}, the optimality condition delivers optimal distancing 
\begin{equation}\label{eq:optimal_distancing}
%\varepsilon_{i}(t) = 1 + \beta \frac{1}{c(t)} I(t) \eta_{i}(t) \text{ or } 
d_i(t) = - \frac{\beta}{c} \eta_{i}(t) I(t),
\end{equation}
provided that the entire distancing path admits an interior solution, i.e., that $d_i(t)\in[0,1]$ for all $t$. One should keep in mind that the marginal value of an increased probability of infection, $\eta_i(t)$, is negative. The extent to which an individual distances is, ceteris paribus, increasing in the infection rate, $\beta$, and the size of the infected population, $I(t)$, and decreasing in the cost parameter, $c$, and the co-state, $\eta_i(t)$. Importantly, the decisions today influence the probability of getting infected both today and in the future, which in turn affects the distancing decisions today\textemdash a fact that is captured by the co-state $\eta_{i}(t)$. 

The current-value co-state variable $\eta_{i}$ follows the adjoint equation
\begin{align}
\dot{\eta}_{i}(t) & = \rho \eta_{i}(t) - \frac{\partial \mathcal{H}_{i}}{\partial p_{i}(t)} \notag \\
& = \eta_{i}(t) \left( \rho + \varepsilon_{i}(t) \beta I(t) \right) + \left( \pi_{S} - \frac{c}{2}(1-\varepsilon_{i}(t))^{2}  - \rho V_{I}\right). \label{eq:eta_dot}
\end{align}
The transversality condition is $\displaystyle \lim_{t \rightarrow \infty} e^{-\rho t}\eta_{i}(t) = 0$. In equilibrium, $\eta = \eta_{i}$ for all $i$. Using the adjoint equation and the transversality condition, we solve for $\eta$. 

\begin{lemma}\label{lemma:eta}
Suppose that the rest of the population is following the strategy $\varepsilon$, and $\varepsilon_{i}$ is the individual $i$'s best response. Then
\begin{equation}\label{eq:eta_i_solved}
\eta_i(t) = - \int_{t}^{\infty} e^{-\rho (s-t)} \frac{1-p_i(s)}{1-p_i(t)} \left( \pi_{S} -  \frac{c}{2}(1-\varepsilon_i(s))^2 - \rho V_{I} \right) ds.
\end{equation}
Let $(S, I, R, \varepsilon, p)$ be an equilibrium. Then
\begin{equation}\label{eq:eta_solved}
\eta(t) = - \int_{t}^{\infty} e^{-\rho (s-t)} \frac{S(s)}{S(t)} \left( \pi_{S} -  \frac{c}{2}(1-\varepsilon(s))^2 - \rho V_{I} \right) ds.
\end{equation} 
\end{lemma}

We term $\pi_{S} - \frac{c}{2}(1-\varepsilon(t))^{2} - \rho V_{I}$ the \emph{susceptibility premium} at time $t$. It is the difference in flow payoffs between being susceptible and being infected. The cost of getting infected, $-\eta(t)$, is the discounted value of the susceptibility premium over time weighted by the conditional probability of being susceptible at each time in the future, $s\geq t$, $\frac{S(s)}{S(t)}$.\footnote{Note that formula (\ref{eq:eta_solved}) extends to more general specifications of the cost of distancing.} The assumption $\pi_S-\frac{c}{2}> \rho V_I$ implies that $\eta$ is negative. In other words, getting infected with certainty is worse than being susceptible and fully distancing: $-\eta(t) >0$. Distancing over a period of time reduces the quality of life and, thus, the susceptibility premium. However, it also decreases the probability that the individual will get infected and rewards her with the premium for a longer period of time. The functional form of $\eta_i$ demonstrates the difficulty of the dynamic problem. Optimal exposure at time $t$ depends on exposure of the remaining individuals through the effect it has on the spread of the infection, as well as on the exposure of individual $i$ at each instance in the future and the effect that this future exposure has on the benefit of getting infected today.

Alternatively, one can decompose $\eta$ in two parts
\begin{align*}
\eta(t) = - \left( V_S(t) - V_I(t) \right)
\end{align*}
where 
\begin{align*}
V_S(t) = \int_{t}^{\infty} e^{-\rho(s-t)} \left( \frac{S(s)}{S(t)} \left( \pi_{S} - \frac{c}{2}(1-\varepsilon(s))^{2} \right) + \left( 1- \frac{S(s)}{S(t)} \right)\rho V_{I} \right) ds
\end{align*}
is the continuation payoff of being susceptible and 
\begin{align*}
V_I(t) = V_I,
\end{align*}
is the continuation payoff of being infected. 

The above discussion implies that characterizing the set of equilibria analytically is untenable. To verify, whether a distancing function $\varepsilon$ can be part of an equilibrium, one needs to posit that the individuals use it, derive $S$, $I$, $R$ and $\eta$, and then verify that $\varepsilon$ is indeed a best reply given the dynamics. This task is made more challenging by the fact that even the SIR model without distancing does not have a tractable closed-form solution and that $\eta$ is pinned down only in the limit rather than at any point.

However, we can make use of the model with an endogenous cost of infection to inform our parameter choices in the constant cost of infection model. The following lemma provides bounds for $\eta$, which enable us to connect the two models. 

\begin{lemma}\label{lemma:eta_bounds}
Let $(S, I, R, \varepsilon, p)$ be an equilibrium. Then
\begin{align}\label{eq:eta_bounds}
-\frac{\pi_S-\rho V_{I}}{\rho} \leq \eta(t) \leq -\frac{\pi_S - \rho V_I-\frac{c}{2}}{\rho+\beta},
\end{align}
and
\begin{equation}
\displaystyle \lim_{t \rightarrow \infty} \eta(t) = - \frac{\pi_{S} - \rho V_{I}}{\rho}. \label{eq:eta_infty}
\end{equation}
If $\dot \eta(0) >0$, then 
\begin{align}
\eta(t) \leq -\frac{\pi_S - \rho V_I-\frac{c}{2}}{\rho}.
\end{align}

\end{lemma}

As time passes, $\eta$ eventually converges to the lower bound\textemdash that is, the cost of getting infected approaches its upper bound. The bound is attained when individuals choose full exposure in perpetuity without facing any risk of becoming infected. This is the scenario in which getting infected would be most costly as there is no need to distance and no risk of future infection. The convergence to this bound is intuitive: as times goes to infinity the infection dies out and obviates the need for distancing. 

The above lemma also provides an upper bound on $\eta$. This bound applies even if $\eta$ is locally decreasing at time $0$. When $\eta$ is increasing at the onset, which occurs if $I_{0}$ is sufficiently small, the upper bound $- \frac{\pi_{S}-\rho V_{I}-\frac{c}{2}}{\rho}$ is approximately tight. This bound corresponds to the cost of infection when individuals are fully distancing from now until eternity.

\begin{figure}[h!]
	\begin{minipage}{0.5\hsize}
		\begin{center}
			\includegraphics[width=\textwidth]{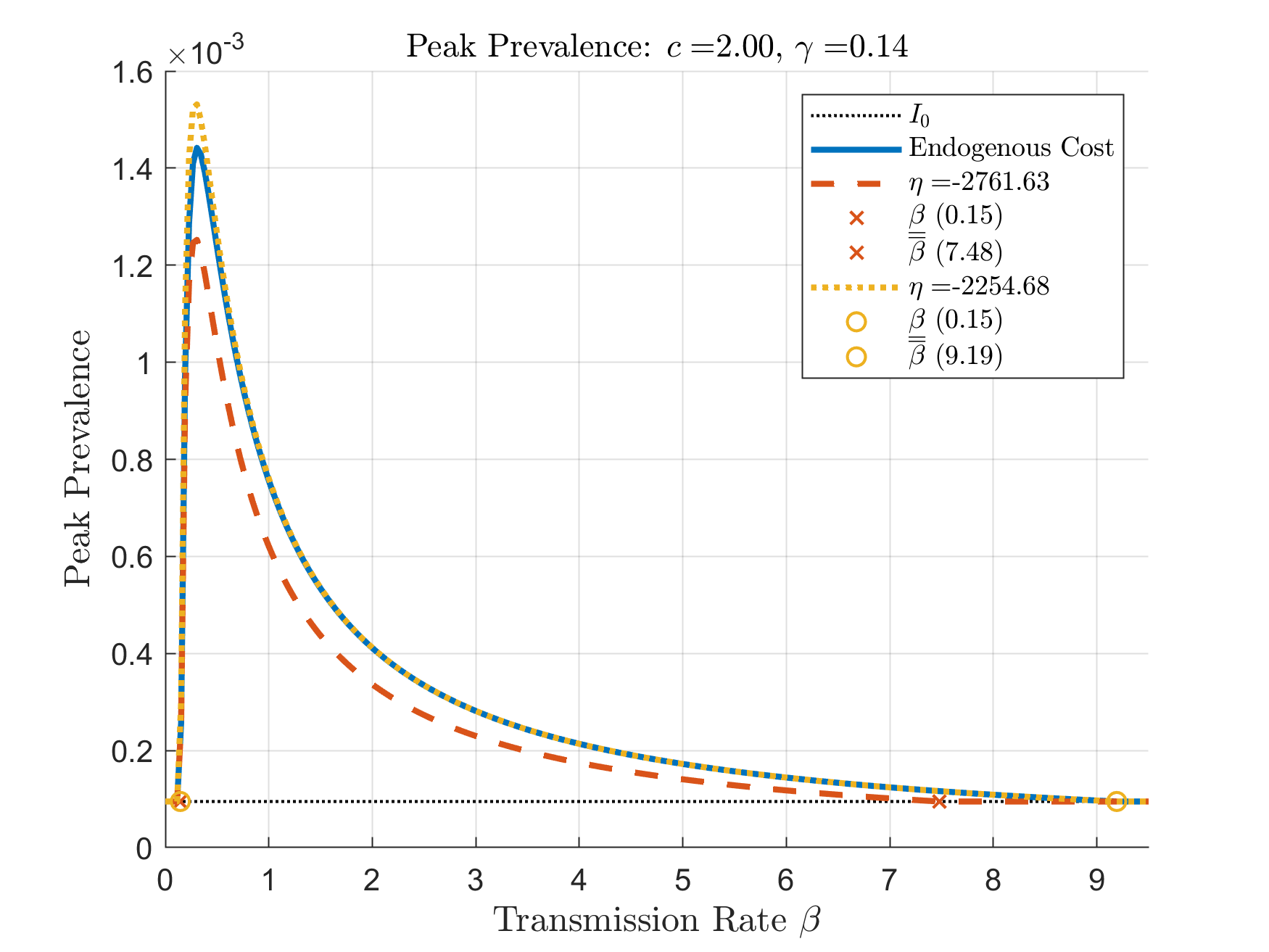}
		\end{center}
		\end{minipage}
		\begin{minipage}{0.5\hsize}
		\begin{center}
			\includegraphics[width=\textwidth]{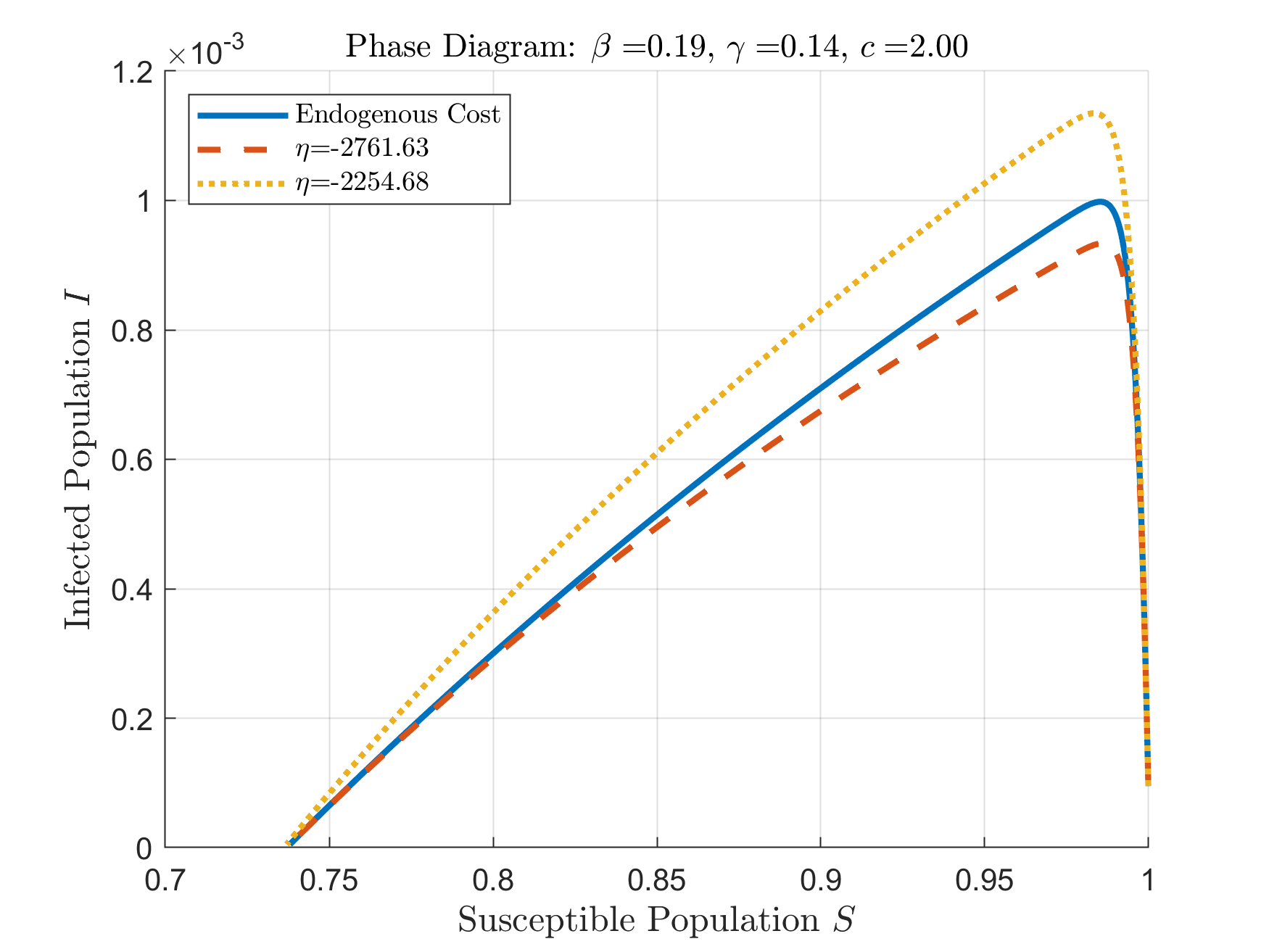}
		\end{center}
	\end{minipage}
	\caption{\emph{Peak Prevalence and Solution Path of the Endogenous Cost of Infection Model.} In the left panel, the solid blue curve shows the peak prevalence in the endogenous cost of infection model as a function of the transmission rate. The dashed and dotted curves reproduce the constant cost of infection model's peak prevalence using the derived bounds on $\eta$. In the right panel, the solid blue curve represents the solution path of the endogenous cost of infection model. The dashed and dotted curves represent the constant cost of infection model's solution paths using the derived bounds on $\eta$.}\label{fig:endogenous_cost}
\end{figure}

Lemma \ref{lemma:eta_bounds} connects the solution paths of the model analyzed here and the model with a fixed cost of infection. Towards that, let $(S,I,R, \varepsilon, p)$ be an equilibrium of the endogenous infection cost model with $\eta$ being the corresponding co-state given by (\ref{eq:eta_solved}). Let $\eta_L$ and $\eta_H$ be the lower and the upper bound on $\eta$ as given by Lemma \ref{lemma:eta_bounds}. Finally, let $(S_j,I_j,R_j,\varepsilon_j)$, for $j \in \{L,H\}$, be the equilibria of the model with the constant cost of infection corresponding to the lower and upper bounds of $\eta$.\\

\begin{prop} \label{prop:phase_comparison_full}
In the phase space, the graph of $(S_{H}, I_{H})$ is above 
that of $(S,I)$, which, in turn, is above that of $(S_{L},I_{L})$.
\end{prop}

We numerically solve the endogenous cost of infection model using commonly used parameters for COVID-19 following \citet{Farboodi_et_al_20} with our objective function. The results for the solution path in the phase diagram are depicted in Figure \ref{fig:endogenous_cost}. First, the curves corresponding to the endogenous cost of infection model are between the curves of the model with the constant cost of infection, with the cost of infection being evaluated at the lower and the upper bound. Importantly, the non-monotonicity of the peak prevalence in $\beta$ is not an artifact of the constant cost of infection model which can be seen in the left panel of the figure.

\section{Behavioral Foundation of Non-linear Contact Rates}
 
The standard SIR model assumes a constant rate of disease transmission which does not incorporate behavioral responses by individuals. As a remedy, SIR models have been proposed in which the transmission rate depends on the current infected population; the pioneering paper is \citet{Capasso_Serio_78}, %; see also \citet{Capasso_93}. 
for an overview of literature see \citet{Funk_et_al_10} and \citet{Verelst_et_al_16}.

Our model provides a behavioral foundation of such SIR models with non-linear contact rates. While the subsequent literature examines a particular functional form of a time-varying contact rate that yields a certain solution path, our modeling can recover the underlying cost of distancing that yields the given functional form. In principle, we can relate the effective disease transmission rate and the cost of distancing. 

In the SIR model of \citet{Capasso_Serio_78} the force of infection $\beta \varepsilon(t) I(t)$ at time $t$ is given by $g(I(t))$, where $g: [0,1] \rightarrow [0, \infty)$ is a non-negative bounded continuously-differentiable function which satisfies the following three assumptions: (i) $g(0) = 0$; (ii) $g'(0)>0$; and (iii) $g(x) \leq g'(0)x$ for all $x \in [0, 1]$. Given this specification, they study the dynamics of the solution path $(S,I)$ qualitatively in the phase space. They also provide a particular solution path by considering a particular function $g$. We will recover the underlying cost of distancing for a general $g$ and the particular $g$ in \citet{Capasso_Serio_78}. 

To that end, with some abuse of notation, denote by $c: [0,1] \rightarrow [0, \infty)$ the cost function of distancing. The cost of distancing at time $t$, therefore, is $c(d(t))$. Assume that $c$ is twice-continuously-differentiable, increasing, convex, and $c'(0)=0$. In this set-up, the first-order condition of problem (\ref{eq:naive_prob}) with respect to distancing is
\begin{align*}
c'(d(t))  = \eta \beta I(t) .
\end{align*}
Since $c$ is strictly convex, $g$ can be recovered as
\begin{equation*}
g(I(t)) = \beta I(t) (1-(c')^{-1}(\eta \beta I(t))).
\end{equation*}
It is easy to verify that $g$ satisfies the three required assumptions. 

\citet{Capasso_Serio_78} consider the particular function $g$ given by
\begin{equation*}
g(I(t))=\frac{\beta I(t)}{1+\frac{I(t)}{\alpha}}.
\end{equation*}
It can be seen that the underlying cost function satisfies
\begin{equation}\label{eq:cost_capasso}
c(d(t)) = -\eta \beta \alpha  ( (1-d(t)) - \log (1-d(t)) ).
\end{equation}
In their specification, thus, the cost of distancing is unbounded when an individual fully distances. This observation allows us to reinterpret the model of \citet{Capasso_Serio_78} as a model of endogenous behavioral responses to a constant infection risk with the cost of distancing function given in equation (\ref{eq:cost_capasso}); in such a model, each individual weights the cost and benefit of distancing at each point in time. Interestingly, \citet{Farboodi_et_al_20} use precisely this cost function in their numerical analysis of the COVID-19 pandemic.\footnote{Note that that they normalize the cost of distancing to $c(0)=0$ so that they have $\alpha =\frac{1}{\eta \beta}$ and an additional but inconsequential $+1$.}

\newpage
\bibliographystyle{ecta}
\bibliography{covid_ref}

\newpage
\appendix

\section{Appendix}\label{sec:appendix}

\begin{proof}[\bf{Proof of Proposition \ref{prop:uniqueness_naive}}]
Any individual's problem (\ref{eq:naive_prob}) is concave; therefore, the first-order condition given by (\ref{eq:naive_distancing}) is also sufficient. This pins down the individual's optimal distancing in the SIR dynamics. Thus, in any equilibrium, if it exists, $(S,I,R)$ is characterized by the system of differential equations $\frac{d}{dt}(S,I,R) = F(t,S,I,R)$, where $F$ is defined by (\ref{eq:naive_dot_S}), (\ref{eq:naive_dot_I}), and (\ref{eq:naive_dot_R}). The initial condition is $(S(0),I(0),R(0)) = (S_{0}, I_{0},0)$. Then, the initial value problem admits a unique solution $(S,I,R)$ on $[0, \infty)$, as the system satisfies the standard conditions. Namely, the function $F$ is continuous on the domain $D=[0, \infty) \times [0,1]^{3}$, and $F$ is uniformly Lipschitz continuous in $(S,I,R)$: there exists a Lipschitz constant $L$ satisfying $\| F(t,S,I,R) - F(t,\tilde{S}, \tilde{I}, \tilde{R}) \| \leq L \| (S, I,R) - (\tilde{S}, \tilde{I}, \tilde{R}) \|$ for each $t \in [0, \infty)$. See, for example, \citet{Walter_98}. Now, $\varepsilon = \varepsilon_{i}$ is uniquely determined, and hence the model admits a unique (symmetric) equilibrium.
\end{proof}

\begin{proof}[{\bf Proof of Proposition \ref{prop:single_peak}}]
Let $\hat{t}$ be as in the supposition of the proposition. We first show $\varepsilon(\hat{t}) \in (0,1)$. Observe that $I(\cdot)$ is always positive as $I$ follows $\dot{I}(t) = I(t)(\beta \varepsilon(t) S(t) - \gamma) \geq - \gamma I(t)$ and $I(0)>0$. Thus, evaluating $\dot I (t)=0$ at $\hat t$ yields $\beta S(\hat{t}) \varepsilon (\hat{t}) = \gamma$. Hence, $\varepsilon(\hat{t}) \in (0,1)$. 

Next, since $\dot{I}$ is differentiable at $\hat{t}$, it follows that $\ddot{I}(\hat{t})$ exists. We show:
\begin{align*}
\ddot{I}(\hat t) & = \beta \left( \dot{S}(\hat t)I(\hat{t})\varepsilon(\hat t) + S(\hat{t})\dot{I}(\hat t)\varepsilon(\hat t) + S(\hat t)I(\hat t)\dot{\varepsilon}(\hat t) \right) - \gamma \dot{I}(\hat t) \\
& = \beta I(\hat t) \left( \dot{S}(\hat t)\varepsilon(\hat t) + S(\hat t)\dot{\varepsilon}(\hat t) \right) \\
& = \beta S(\hat t)I(\hat t) \left( - \beta I(\hat t)\varepsilon^{2}(\hat t) + \dot{\varepsilon}(\hat t) \right) \\
& =  -\beta S(\hat t)I^2(\hat t)\varepsilon^{2}(\hat t) < 0.
\end{align*}
The second equality follows from $\dot I(\hat t) = 0$, the third from (\ref{eq:S_dot}), and the forth from 
\begin{align}\label{eq:epsilon_dot_peak}
\dot{\varepsilon}(\hat t) &= \frac{\eta \beta}{c} \dot{I}(\hat t) = 0,
\end{align}
which, in turn, follows from optimality condition (\ref{eq:naive_distancing}) and $\varepsilon(\hat{t}) \in (0,1)$. Equation (\ref{eq:epsilon_dot_peak}) also yields the second assertion.
\end{proof}
 
\begin{proof}[{\bf Proof of Proposition \ref{prop:I_0}}]
The first statement follows from the equality $\dot I(0) = \frac{I_0}{\gamma} \left( R^b_0 -1 \right)$. 

Part (\ref{itm:hurlyburly1}): From (\ref{eq:naive_dot_I}) it follows that 
\begin{align*}
\dot I(0) >0 \text{ if and only if } I_0 \left( \beta \left( 1+ \frac{\beta \eta}{c} I_{0}\right)(1-I_0) - \gamma \right) >0.
\end{align*}
Therefore, $\dot I(0) > 0$ if and only if $\beta \in (\underline \beta, \overline \beta)$ where $\underline \beta$ and $\overline \beta$ are solutions to the quadratic equation
\begin{align}\label{eq:I_dot_0_quadratic}
\beta \left( 1+ \frac{\eta I_{0}}{c} \beta \right)(1-I_{0}) - \gamma =0.
\end{align}
It can be seen that the discriminant of the quadratic equation is positive if and only if $I_{0} < \frac{1}{1-\frac{4 \eta \gamma}{c}}$. Note that the solid curve in the left panel of Figure \ref{fig:I_dot_0_naive} corresponds to equation (\ref{eq:I_dot_0_quadratic}). Since $\varepsilon(0) = 1+ \frac{\eta I_{0}}{c} \beta <1$ and $I_{0} >0$, the left-hand side of the above equation is negative at $\beta = \frac{\gamma}{1-I_{0}}$. Thus, $\underline{\beta} > \frac{\gamma}{1-I_{0}}$. If $\beta = - \frac{c}{\eta I_{0}}$, then $\varepsilon(0)=0$ and $\dot{I}(0)<0$. Thus, $\overline{\beta} < - \frac{c}{\eta I_{0}}$.

For the second assertion, let $\beta \in (\underline{\beta}, \overline{\beta})$. Then, it follows from the previous arguments that $\varepsilon(0)>0$ and $\dot{I}(0)>0$. Suppose to the contrary that $1 + \frac{\eta \beta}{c} I(t)=0$ at some $t$. Without loss, assume that $1 + \frac{\eta \beta}{c} I(s)>0$ for all $s \in [0, t)$. At $t$, $\varepsilon(t)=0$, $\dot{S}(t)=0$, and $\dot{I}(t) = - \gamma I(t) <0$. Then, there exists $\hat{s} \in (0, t)$ such that $\dot{I}(\hat{s})=0$, i.e., $S(\hat{s}) \varepsilon(\hat{s}) = \frac{\gamma}{\beta}$. Thus, before $I$ hits $\frac{c}{\beta (-\eta)}$ at $t$, $(S,I)$ has to satisfy $\dot{I}(\hat{s})=0$ and $\dot{S}(\hat{s})<0$. That is, at $\hat{s}$, $I$ peaks. Then, $0 = 1 + \frac{\eta \beta}{c} I(t)>1 + \frac{\eta \beta}{c} I(\hat{s})>0$, a contradiction. \newline

\noindent Part (\ref{itm:hurlyburly2}): Let $I_{0} \geq \frac{1}{1-\frac{4 \eta \gamma}{c}}$. Then, the quadratic equation (\ref{eq:I_dot_0_quadratic}) has at most one solution. Thus, $\dot{I}(0) \leq 0$. Proposition \ref{prop:single_peak} then implies that if $\dot{I}(t) \leq 0$ for some $t$ (take $t=0$) then $\dot{I}(s) <0$ for all $s >t$.  \newline
\end{proof}

\begin{proof}[{\bf Proof of Proposition \ref{prop:phase}}]
By Assumption $d(0)<1$, it can be seen that $\varepsilon(t) \in (0,1)$ for all $t$. Then, we have
\begin{align*}
\frac{dS}{d(S+I)} &= \frac{\beta}{\gamma} S \varepsilon = \frac{\beta}{\gamma} \left( - \frac{\beta \eta}{c} S^{2} + S + \frac{\beta \eta}{c} (S+I) S \right),
\end{align*}
where the first equality follows from dividing (\ref{eq:S_dot}) by the sum of (\ref{eq:S_dot}) and (\ref{eq:I_dot}), and the second uses (\ref{eq:naive_distancing}) and simple manipulations.
The above expression can be rewritten as
\begin{equation*}
\frac{d}{d(S+I)} \left( \frac{1}{S} \right) + \left( \frac{\beta}{\gamma} + \frac{\beta^{2}\eta}{\gamma c} (S+I) \right) \frac{1}{S} = \frac{\beta^{2}\eta}{\gamma c},
\end{equation*}
which is a linear first-order differential equation with respect to $\frac{1}{S}$ and $(S+I)$. For ease of notation, let $y = \frac{1}{S}$ and $x = S+I$. Then, 
\begin{equation}\label{eq:phase_dif_eq_naive_xy_formulate}
\frac{dy}{dx} + \left( \frac{\beta}{\gamma} + \frac{\beta^{2}\eta}{\gamma c} x \right) y = \frac{\beta^{2}\eta}{\gamma c}.
\end{equation}
Let $\mu(x) := \exp \left( \int \left( \frac{\beta}{\gamma} + \frac{\beta^{2}\eta}{\gamma c} x \right) dx \right)$ be the integrating factor. We have
\begin{equation}\label{eq:magic}
\mu(x) = k \cdot \exp \left( \frac{\beta^{2}\eta}{2\gamma c} \left( x + \frac{c}{\beta \eta} \right)^{2} \right),
\end{equation}
where $k$ is the constant of integration. Then, equation (\ref{eq:phase_dif_eq_naive_xy_formulate}) reduces to
\begin{align}
\frac{d}{dx} \left[ \mu(x) y \right] & = \mu(x) \left[ \frac{d}{dx} y + \left( \frac{\beta}{\gamma} + \frac{\beta^{2}\eta}{\gamma c} x \right) y \right] = \mu(x) \frac{\beta^{2}\eta}{\gamma c}. \label{eq:phase_dif_eq_naive_xy}
\end{align}
Integrating the outer most sides of Expression (\ref{eq:phase_dif_eq_naive_xy}) and using (\ref{eq:magic}) yield
\begin{equation}\label{eq:phase_dif_eq_naive_integrated}
\left[ \exp \left( \frac{\beta^{2}\eta}{2\gamma c} \left( x + \frac{c}{\beta \eta} \right)^{2} \right) y \right]_{S+I}^{1} = \frac{\beta^{2}\eta}{\gamma c} \int_{S+I}^{1} \exp \left( \frac{\beta^{2}\eta}{2\gamma c} \left( x + \frac{c}{\beta \eta} \right)^{2} \right) dx.
\end{equation}
The left-hand side of (\ref{eq:phase_dif_eq_naive_integrated}) reduces to
\begin{equation*}
\exp \left( \frac{\beta^{2}\eta}{2\gamma c} \left( 1 + \frac{c}{\beta \eta} \right)^{2} \right) \frac{1}{S_{0}} - \exp \left( \frac{\beta^{2}\eta}{2\gamma c} \left( S+I + \frac{c}{\beta \eta} \right)^{2} \right) \frac{1}{S}.
\end{equation*}
For the right-hand side of (\ref{eq:phase_dif_eq_naive_integrated}), let $v = \beta \sqrt{\frac{-\eta}{2\gamma c}} \left( x + \frac{c}{\beta \eta}\right)$. Since $\frac{dv}{dx} = \beta \sqrt{\frac{-\eta}{2\gamma c}}$, the right-hand side of (\ref{eq:phase_dif_eq_naive_integrated}) reduces to
\begin{align*}
 - \beta \sqrt{\frac{2(-\eta)}{\gamma c}} \int_{\beta \sqrt{\frac{-\eta}{2\gamma c}} \left( S+I + \frac{c}{\beta \eta}\right)}^{\beta \sqrt{\frac{-\eta}{2\gamma c}} \left( 1 + \frac{c}{\beta \eta}\right)} e^{-v^2} dv.
\end{align*}
Hence, we can rewrite equation (\ref{eq:phase_dif_eq_naive_integrated}) as 
\begin{align*}
\exp \left( \frac{\beta^{2}\eta}{2\gamma c} \left( S+I + \frac{c}{\beta \eta} \right)^{2} \right) \frac{1}{S} = \exp \left( \frac{\beta^{2}\eta}{2\gamma c} \left( 1 + \frac{c}{\beta \eta} \right)^{2} \right) \frac{1}{S_{0}} + \beta \sqrt{\frac{2(-\eta)}{\gamma c}} \int_{\beta \sqrt{\frac{-\eta}{2\gamma c}} \left( S+I + \frac{c}{\beta \eta}\right)}^{\beta \sqrt{\frac{-\eta}{2\gamma c}} \left( 1 + \frac{c}{\beta \eta}\right)} e^{-v^2} dv,
\end{align*}
and finally we obtain (\ref{eq:phase}), as desired.
%\begin{align*}
%S = \frac{\exp \left( \frac{\beta^{2}\eta}{2\gamma c} \left( S+I + \frac{c}{\beta \eta} \right)^{2} \right)}{\displaystyle \exp \left( \frac{\beta^{2}\eta}{2\gamma c} \left( 1 + \frac{c}{\beta \eta} \right)^{2} \right) \frac{1}{S_{0}} + 2\beta \sqrt{\frac{(-\eta)}{2\gamma c}} \int_{\beta \sqrt{\frac{-\eta}{2\gamma c}} \left( S+I + \frac{c}{\beta \eta}\right)}^{\beta \sqrt{\frac{-\eta}{2\gamma c}} \left( 1 + \frac{c}{\beta \eta}\right)} e^{-t^2} dt} .
%\end{align*}
\end{proof}

\begin{proof}[{\bf Proof of Proposition \ref{prop:phase_comparison}}]
\noindent Part (\ref{itm:prop_phase_comparison_1}): We prove the assertion with respect to $\frac{\eta}{c}$. Denote by a point $(S, I(S))$ on the solution path. Differentiating the quotient differential equation $\frac{dI}{dS}$ with respect to $\frac{\eta}{c}$ at a fixed point $(S, I(S))$ yields
\begin{align*}
\frac{\partial}{\partial \frac{\eta}{c}}\left( \frac{dI}{dS}\left( \frac{\eta}{c} \right)\right)=-\frac{\gamma I}{S}\frac{1}{(1+\beta I \frac{\eta}{c})^2}<0.
\end{align*}
Now, take $\frac{\eta}{c}$ and $\frac{\tilde{\eta}}{\tilde{c}}$ with $\frac{\eta}{c} < \frac{\tilde{\eta}}{\tilde{c}}$. Denote by $(\tilde{S},\tilde{I})$ the solution path associated with $\tilde{\eta}$ and $\tilde{c}$. Since $\frac{dI}{dS}<0$ at $(S_{0},I_{0})$, it follows that $\tilde{I}(S_{0}-\delta)>I(S_{0}-\delta)$ for some small $\delta>0$. Now, it is sufficient to show that two curves $I$ and $\tilde{I}$ do not intersect. Suppose to the contrary that $I$ and $\tilde{I}$ did intersect. Then, for $\overline{S} := \sup \{ S \in (0,S_{0}) \mid \tilde{I}(S)=I(S) \}$, it would have to be the case that $\frac{d\tilde{I}}{d\tilde{S}}(\overline{S})>\frac{dI}{dS}(\overline{S})$. However, this cannot happen as $\frac{d}{d\frac{\eta}{c}}\left(\frac{dI}{dS}(S)\right)<0$ at any point $(S,I(S))$.

\noindent Part (\ref{itm:prop_phase_comparison_2}): We first show $S(t) \geq \hat{S}(t)$ for all $t\geq0$. Suppose to the contrary that there exists some $\tilde t$ such that $S(\tilde t) < \hat{S}(\tilde t)$. At time $0$, $S(0) = \hat{S}(0)$ and $\dot{S}(0) > \dot{\hat{S}}(0)$. Thus, there exists an interval in which $S(\cdot) > \hat{S}(\cdot)$. Then there would have to exist $t_{0}$ such that $S(t_{0}) = \hat{S}(t_{0})$ and $\dot{S}(t_{0}) < \dot{\hat{S}}(t_{0})$. However, it follows from $S(t_{0}) = \hat{S}(t_{0})$ and the previous argument that $I(t_{0}) \leq \hat{I}(t_{0})$, and thus 
\begin{equation*}
\dot{S}(t_{0}) = - \beta \varepsilon(t_{0}) S(t_{0}) I(t_{0}) > - \beta S(t_{0}) I(t_{0}) > - \beta \hat{S}(t_{0}) \hat{I}(t_{0}) =  \dot{\hat{S}}(t_{0}),
\end{equation*}
which is impossible.

Next, we show $R(\cdot) \leq \hat{R}(\cdot)$. Suppose to the contrary that $R(t_{0}) > \hat{R}(t_{0})$ for some $t_{0} \in (0, \infty)$. Define $t_{1}:=\sup \{ t \in [0,t_{0}] \mid R(t) = \hat{R}(t) \text{ and } R(s) > \hat{R}(s) \text{ for all } s \in (t, t_{0}] \}$.\footnote{If $R(t) > \hat{R}(t)$ for all $t \in (0, t_{0}]$ then $t_{1}=0$. If not, $t_{1}=\sup \{ t \in [0, t_{0}] \mid R(t) = \hat{R}(t) \}$.} Since $R(t_{1}) = \hat{R}(t_{1})$, it follows that $S(t_{1})+I(t_{1}) = \hat{S}(t_{1}) + \hat{I}(t_{1})$. There exists a small $\delta \in (0, t_{0}-t_{1})$ such that $\hat{I}(t) < I(t)$ for all $t \in (t_{1}, t_{1}+\delta)$ because $R(t)-R(t_1)=\gamma \int_{t_1}^t I(s) ds$, $\hat{R}(t)-\hat{R}(t_1)=\gamma \int_{t_1}^t \hat{I}(s) ds$, and $R(t_1)=\hat{R}(t_1)$. Also, it follows from the previous argument that $\hat{S}(t) \leq S(t)$ for all $t \in (t_{1}, t_{1}+\delta)$. Thus, $\hat{R}(t) > R(t)$ for all $t \in (t_{1}, t_{1}+\delta)$, a contradiction.
\end{proof}

\begin{proof}[{\bf Proof of Proposition \ref{prop:I_peak}}] 
We prove the result with respect to $c$ first, then with respect to $\beta$. \\
Part (\ref{itm:prop_I_peak_2}): The proof of Proposition \ref{prop:I_0} has established that 
\begin{align*}
\dot I(0) >0 \text{ if and only if } I_0 \left( \beta \left( 1+ \frac{\beta \eta}{c} I_{0}\right)(1-I_0) - \gamma \right) >0.
\end{align*}
Therefore, $\dot I(0) > 0$ if and only if $c > \underline c$, where $\underline c$ can be recovered from the above inequality; remember that $\eta < 0$. The peak prevalence when $c \leq \bar c$ is $I_0$. If $c> \underline{c}$, then the peak prevalence is strictly greater than $I_{0}$. We show that the peak prevalence is strictly increasing in $c >\underline{c}$. Differentiating (\ref{eq:phase_diff_eq_naive}) with respect to $c$, while holding $S$ and $I$ fixed, yields\footnote{Recall that when $\dot{I}(0)>0$, exposure is interior for all $t$.}
\begin{align*}
\frac{\partial}{\partial c} \left( \frac{dI}{dS}\right)& = \frac{\gamma \eta I}{c^2 S \left( 1+ \beta I \frac{\eta}{c}\right)^2} < 0,
\end{align*}
where the inequality is due to $\eta <0$. If two solution paths corresponding to $c$ and $c'>c$ intersect at some point, the solution path corresponding to $c'$ has a smaller slope. A certain point of intersection is the beginning of the infection $(S_0, I_0)$. At this point in the graph with $S$ on the horizontal and $I$ on the vertical axis the solution path corresponding to $c'$ is steeper; the solution paths are decreasing at $(S_0, I_0)$. Just below $S_0$, then, the solution path corresponding to $c'$ is above the one corresponding to $c$. If they were to intersect at some other $S < S_0$, the solution path corresponding to $c'$ would have to intersect the solution path corresponding to $c$ from above and stay below it. This would contradict the finding that the solution path corresponding to $c'$ is above the one corresponding to $c$ for $S$ slightly below $S_0$. Finally, given that the solution path under $c'$ is above the solution path under $c$, the peak of infection under $c'$ must be higher than the peak of infection under $c$.

\noindent Part (\ref{itm:prop_I_peak_1}): We break up the proof for $\beta$ into two steps.

Step 1: $I^{\ast}$ is decreasing in $\beta$ for $\beta \in [-\frac{c}{2I_0 \eta}, \overline{\beta}]$; notice that $\eta$ is negative, thus, the lower bound is positive. The derivative of the quotient differential equation (\ref{eq:phase_diff_eq_naive}) at a given point $(S,I(S))$ with respect to $\beta$ is
\begin{align}\label{eq:db dS}
\frac{\partial}{\partial \beta}\left( \frac{dI}{dS}(\beta) \right) &= - \frac{\gamma}{\beta^2 S}\frac{1+2 \frac{\beta \eta}{c}I(S)}{(1+\frac{\beta \eta}{c}I(S))^2}.	
\end{align}
The above derivative evaluated at $(S_0,I_0)$ is greater or equal to $0$, for $\beta \geq \frac{c}{-2I_0 \eta}$. This means that at $(S_0,I_0)$, a higher $\beta$ leads to a slower spread of the infection when the starting $\beta$ is high enough. At $(S_0,I_0)$ solution paths are decreasing, thus the positive derivative with respect to $\beta$ means that the solution path becomes flatter as $\beta$ increases. That is, around $(S_0,I_0)$ the solution path corresponding to a higher $\beta$ is, therefore, below the one with the lower beta.

Moreover, $\frac{\partial}{\partial \beta}\left( \frac{dI}{dS}(\beta) \right) \geq 0$ at $(S_0, I_0)$, for $\beta \geq \frac{c}{-2I_0 \eta}$, implies that the same is true for all $(S,I)$ with $I > I_0$. This means that if two solution paths corresponding to some $\beta$ and $\beta'>\beta$ in $[-\frac{c}{2I_0 \eta}, \overline \beta]$ intersect, then the solution path corresponding to $\beta'$ must have a larger slope. One such point of intersection is $(S_0, I_0)$. Therefore, a solution path for $\beta'$ is below the one of $\beta$ just below $S_0$ and it cannot intersect it anymore as long as $I \geq I_0$. In other words, the solution path of $\beta'$ is strictly below the solution path of $\beta$ for all $I > I_0$. The maximum of $I$ for $\beta'$ is, therefore, strictly below the maximum of $I$ for $\beta$. 

Step 2: There exists a $\beta_1$ such that $I^{\ast}$ is increasing in $\beta$ on $(\underline \beta, \beta_1)$. We divide this step into three substeps. First, we show that the peak $I^{\ast}$ is continuous in $\beta$. Then, we show that $\frac{\partial}{\partial\beta}\left( \frac{dI}{dS}(\underline{\beta}) \right)<0$ along the entire solution path. Finally, we combine these two insights to show that for $\beta>\underline{\beta}$ but sufficiently close $\frac{\partial}{\partial\beta}\left( \frac{dI}{dS}(\beta) \right)<0$ implying that the peak is increasing in $\beta$ for $\beta\in(\underline{\beta},\underline{\beta}+\delta)$ for some $\delta>0$.

Step 2.1: We argue that $I^{\ast}$ is continuous in $\beta\in(0,\overline{\beta})$. For $\beta\in(0, \underline{\beta})$, $I^\ast=I_0$. 

For $\beta \in(\underline{\beta},\overline{\beta})$, in the $(S,I)$-phase space, $(S, I) = (S^{\ast}, I^{\ast})$ satisfies (\ref{eq:phase}) and $\frac{dI}{dS}=0$, i.e., $S^{\ast} = \frac{\gamma}{\beta} \frac{1}{1+\frac{\eta \beta}{c}I^{\ast}}$. Substituting the latter equation into the former and rearranging, we obtain
\begin{align*}
& \exp \left( \frac{\eta}{2\gamma c} \left( \beta + \frac{c}{\eta} \right)^{2} \right) \frac{1}{S_{0}} + \beta \sqrt{\frac{(-2\eta)}{\gamma c}} \int_{\sqrt{\frac{-\eta}{2\gamma c}} \left( \frac{\gamma}{1+\frac{\beta \eta}{c}I^{\ast}}+\beta I^{\ast} + \frac{c}{\eta}\right)}^{\sqrt{\frac{-\eta}{2\gamma c}} \left( \beta + \frac{c}{\eta}\right)} e^{-v^2} dv \notag \\
= & \frac{1}{\gamma} \left( \beta+\frac{\beta^{2} \eta}{c}I^{\ast} \right) \exp \left( \frac{\eta}{2\gamma c} \left( \frac{\gamma}{1+\frac{\beta \eta}{c}I^{\ast}}+\beta I^{\ast} + \frac{c}{\eta} \right)^{2} \right). %\label{eq:I_peak_SI}
\end{align*}
This implies that $I^{\ast}$ is differentiable and thus continuous in $\beta$ for $\beta\in(\underline{\beta},\overline{\beta})$.

Finally, if $\beta=\underline{\beta}$ then $(S^\ast,I^\ast)=(S_0,I_0)$ satisfies the above implicit equation. Thus, $I^\ast$ is also continuous at $\beta=\underline{\beta}$.

Step 2.2: $\frac{\partial}{\partial\beta}\left( \frac{dI}{dS}(\beta) \right)<0$, along the entire solution path whenever it holds along the path that $1+2 \frac{\underline \beta\eta}{c}I(S) > 0$. This is satisfied for $\beta=\underline{\beta}$.

Recall that the sign of $\frac{\partial}{\partial\beta}\left( \frac{dI}{dS}(\beta) \right)$ at each $(S,I(S))$ is determined by the negative of the sign of $1+2\frac{\beta \eta}{c}I(S)$. Thus, it is sufficient for the derivative to be negative along the entire path that $1+2\frac{\beta \eta}{c}I^\ast>0$. 

Observe that the solution $\underline{\beta}$ of equation (\ref{eq:I_dot_0_quadratic}) is given by
\begin{equation*}
\underline \beta = -\frac{c}{2\eta I_0}\left( 1- \left(1+4\frac{\eta\gamma}{c}\frac{I_0}{S_0}\right)^{\frac{1}{2}}\right).
\end{equation*}
Therefore,
\begin{align*}
1+2 \frac{\underline \beta\eta}{c}I_0 &= \left(1+4\frac{\eta\gamma}{c}\frac{I_0}{S_0}\right)^{\frac{1}{2}} >0,
\end{align*}
where the inequality follows due to the assumption on $I_0$ in the statement of the result. Consequently,
\begin{align*}
\frac{\partial}{\partial \beta}\left( \frac{dI}{dS}(\underline \beta) \right) <0.
\end{align*}

Step 2.3: $I^{\ast}$ is increasing in $\beta$ for $\beta \in (\underline \beta, \underline \beta + \delta)$ for some $\delta>0$. 

Since $1+2 \frac{\underline \beta\eta}{c}I_0 > 0$, there exists a $\delta_1 >0$ such that $1+2 \frac{ \beta\eta}{c}I_0 > 0$ for all $\beta \in [\underline \beta, \underline \beta + \delta_1)$.

By continuity of the peak, for every $\delta_2>0$, there exists a $\delta_3 >0$, such that $\beta \in [\underline \beta, \underline \beta + \delta_3)$ implies $I^{\ast}(\beta) < I_0+\delta_2$. Choose $\delta_2$ to correspond to the $\delta_1$ argued above Step 2.2, and let $\delta_3$ corresponds to such $\delta_2$. This guarantees that we consider $\beta$ to lie in a range such that the peak is sufficiently low to ensure that the slopes of the solution paths can be ordered by comparing $\beta$.

By Steps 2.1 and 2.2, for any such $\beta$, $1+2 \frac{ \beta\eta}{c}I(S)>0$  and therefore  $\frac{\partial}{\partial \beta}\left( \frac{dI}{dS}( \beta) \right) <0$. This implies that whenever two solution paths corresponding to different $\beta$ in $(\underline \beta, \underline \beta + \delta_3)$ intersect at a point, the one with the higher $\beta$ has the smaller slope. Indeed, one point of intersection is $(S_0, I_0)$. The solution path with a higher $\beta$ must be steeper than the other path; the two are decreasing at $(S_0, I_0)$. Suppose the two solution paths were to intersect at some $S<S_0$ and let $\tilde S$ be the largest such $S$. Then due to $\frac{\partial}{\partial \beta}\left( \frac{dI}{dS}( \beta) \right) <0$, the solution path with the higher $\beta$ would need to intersect the solution path with the lower beta from above, and fall below it for $S>\tilde S$. But this contradicts the fact that at $S_0$ the solution path corresponding to a higher $\beta$ is above the one with the lower beta.
\end{proof}

\begin{proof}[{\bf Proof of Proposition \ref{prop:epidemic size}}]
First, we show $S_{0} e^{- \frac{\beta}{\gamma}} \leq \hat{S}_{\infty}$. 
Let $\displaystyle \hat{R}_{\infty} := \lim_{t \rightarrow \infty} \hat{R}(t)$. For the purpose of illustration, let $\hat{\varepsilon}(\cdot)=1$. It follows from the SIR dynamics that 
\begin{align*}
\hat{S}_{\infty} & = S_{0} \exp \left( - \beta \int_{0}^{\infty} \hat{\varepsilon}(s) \hat{I}(s)ds \right) \geq S_{0} \exp \left( - \beta \int_{0}^{\infty} \hat{I}(s)ds \right) = S_{0} \exp \left( - \frac{\beta}{\gamma} \hat{R}_{\infty} \right) \geq S_{0} e^{- \frac{\beta}{\gamma}}.
\end{align*}
The first equality follows in the same way as (\ref{eq:S_ratio}) is derived, by letting $t_{0}=t$ and $t_{1}=\infty$. The first inequality follows because $\hat{\varepsilon}(\cdot) \leq 1$. Indeed, it follows with equality 
in the standard SIR model $\hat{\varepsilon}(\cdot) =1$. The second equality follows from integrating (\ref{eq:R_dot}) (precisely, with $R$ and $I$ replaced by $\hat{R}$ and $\hat{I}$, respectively). The second inequality follows because $\hat{R}_{\infty} \leq 1$. 

Second, we show $\hat{S}_{\infty} \leq S_{\infty}$. Since $\hat{S}(t) \leq S(t)$ holds by Proposition \ref{prop:phase_comparison} (\ref{itm:prop_phase_comparison_2}), letting $t \rightarrow \infty$ yields the desired result.

Third, we show $S_{\infty} < \frac{\gamma}{\beta}$ in two steps. The first step establishes $S_{\infty} \leq \frac{\gamma}{\beta}$. Suppose not. As $S(t)$ is weakly decreasing throughout, there exists a $\delta>0$ such that %$t_{0} \in [0, \infty)$ such that 
$S(t) \geq \delta + \frac{\gamma}{\beta}$ for all $t \geq 0$. Since $\displaystyle \lim_{t \rightarrow \infty} \varepsilon (t)=1$ and $\delta>0$, for a given $\kappa \in (0, \delta)$, there exists $t_{1} \in [t_{0},\infty)$ such that $\delta \varepsilon(t) - \frac{\gamma}{\beta}(1-\varepsilon(t)) > \kappa$ for all $t \geq t_{1}$. Then, for all $t \geq t_{1}$, we have 
\begin{align*}
\dot{I}(t) & = \beta I(t) (S(t)\varepsilon(t) - \frac{\gamma}{\beta})  \geq \beta I(t) ( (\delta + \frac{\gamma}{\beta}) \varepsilon(t) - \frac{\gamma}{\beta} )  > \beta I(t) \kappa,
\end{align*}
that is, $\frac{\dot{I}(t)}{I(t)} > \beta \kappa$ (note that, since $\dot{I}(t) \geq - \gamma I(t)$, $I(t)$ is always positive: $I(t) \geq I(0)e^{-\gamma t} >0$). Thus, $I(t) \geq I(t_{1}) e^{\beta \kappa t}$, which yields $I_{\infty} = + \infty$. This is a contradiction to $I_{\infty}=0$.

The second step establishes $S_{\infty} \neq \frac{\gamma}{\beta}$. Suppose to the contrary $S_{\infty} = \frac{\gamma}{\beta}$. Then, $\frac{dI}{dS}(S_\infty) = -1 + \frac{\gamma}{\beta} \frac{1}{S_{\infty}} = 0$ as $\displaystyle\lim_{t\rightarrow \infty}\varepsilon(t)=1$. However, note that 
\begin{align*}
\frac{d}{dS}\frac{dI}{dS}(S_\infty)&=-\frac{\gamma}{\beta}\frac{1}{\varepsilon(I(S)) S}\left(\frac{1}{S}+\frac{1}{\varepsilon(I(S))}\frac{d\varepsilon(I(S))}{dI(S)}\frac{dI}{dS}\right)\\
&=-\frac{\gamma}{\beta}\frac{1}{\varepsilon(I(S)) S^2}<0
\end{align*}
as $\frac{dI}{dS}(S_\infty)=0$, where $\varepsilon(I(S))=1+\beta \frac{\eta}{c} I(S)$. Thus, there is a $\delta>0$ such that for $S\in (S_\infty,S_\infty+\delta)$, $\frac{dI}{dS}(S_\infty+\delta)<0$ and, hence, that $I(S_\infty+\delta)<0$, a contradiction. Thus, $S_\infty<\frac{\gamma}{\beta}$.
\end{proof}

\begin{proof}[\bf{Proof of Proposition \ref{prop:S_infty_monotone}}]
It follows from Proposition \ref{prop:phase_comparison} that $S_{\infty}$ is increasing in $c$. Thus, we show that $S_{\infty}$ is decreasing in $\beta$ for the following three cases: (1) $\beta \in [0, \underline{\beta}]$; (2) $\beta \in [\underline{\beta}, \overline{\beta}]$; and (3) $\beta \in [\overline{\beta}, \frac{c}{(-\eta)I_{0}}]$. \newline

\noindent \text{Case 1.} Let $\beta \in [0, \underline{\beta}]$. In this case, $I^{\ast} = I_{0}$, and $\dot{I}(t) <0$ for all $t \in (0, \infty)$. Also, $\varepsilon(\cdot) \in (0,1)$ as $\varepsilon(t)=1+\beta \frac{\eta}{c}I(t)$ is decreasing in $I(t)$, $I(t)>0$ is decreasing, and $\beta<-\frac{c}{\eta I(0)}$.

The derivative of the quotient differential equation with respect to $\beta$ at $(S,I(S))$ is
\begin{equation*}
\frac{\partial}{\partial \beta} \frac{dI}{dS} = - \frac{\gamma}{\beta S}\frac{1}{\beta(1+\frac{\beta \eta}{c}I(S))}\left(1+2 \frac{\beta \eta}{c}I(S)\right) <0.	
\end{equation*}
This implies that, for any $\beta,\beta'\in[0, \underline{\beta}]$ with $\beta < \beta'$, the solution path associated with $\beta'$ has a flatter slope than the one associated with $\beta$ at any point $S \in (S_{0}, S_{\infty}(\beta))$, where $S_{\infty}(\beta)$ is $S_{\infty}$ associated with $\beta$. Thus, $I(S_{\infty}(\beta))>0$ for the solution path associated with $\beta'$, and hence $S_{\infty}(\beta') < S_{\infty}(\beta)$. \newline

\noindent \text{Case 2.} Let $\beta \in [\underline{\beta}, \overline{\beta}]$. In this case, $\dot{I}(0) \geq 0$ and $\varepsilon(\cdot) \in (0,1)$. Substituting $(S,I)=(S_{\infty},0)$ into (\ref{eq:phase}) yields 
\begin{align}\label{eq:S_infty_naive}
S_{\infty} = \frac{\exp \left( \frac{\eta}{2\gamma c} \left( \beta S_{\infty} + \frac{c}{\eta} \right)^{2} \right)}{\displaystyle \exp \left( \frac{\eta}{2\gamma c} \left( \beta + \frac{c}{\eta} \right)^{2} \right) \frac{1}{S_{0}} + 2\beta \sqrt{\frac{(-\eta)}{2\gamma c}} \int_{\sqrt{\frac{-\eta}{2\gamma c}} \left( \beta S_{\infty} + \frac{c}{\eta}\right)}^{\sqrt{\frac{-\eta}{2\gamma c}} \left( \beta + \frac{c}{\eta}\right)} e^{-v^2} dv} .
\end{align}
Rewriting Expression (\ref{eq:S_infty_naive}), 
\begin{equation}\label{eq:S_infty_naive_reformulate}
\exp \left( \frac{\eta}{2\gamma c} \left( \beta + \frac{c}{\eta} \right)^{2} \right) \frac{S_{\infty}}{S_{0}} + 2\beta S_{\infty} \sqrt{\frac{(-\eta)}{2\gamma c}} \int_{\sqrt{\frac{-\eta}{2\gamma c}} \left( \beta S_{\infty} + \frac{c}{\eta}\right)}^{\sqrt{\frac{-\eta}{2\gamma c}} \left( \beta + \frac{c}{\eta}\right)} e^{-v^2} dv
= \exp \left( \frac{\eta}{2\gamma c} \left( \beta S_{\infty} + \frac{c}{\eta} \right)^{2} \right).
\end{equation}
For the right-hand side, 
\begin{align*}
\frac{\partial}{\partial \beta} (\mathrm{RHS}) & = \underbrace{\exp \left( \frac{\eta}{2\gamma c} \left( \beta S_{\infty} + \frac{c}{\eta} \right)^{2} \right)}_{=(\mathrm{RHS})} \frac{\eta}{\gamma c} \left( \beta S_{\infty} + \frac{c}{\eta} \right) \left( S_{\infty} + \beta \frac{\partial S_{\infty}}{\partial \beta} \right) .
\end{align*}
For the left-hand side, we obtain 
\begin{align*}
\frac{\partial}{\partial \beta} (\mathrm{LHS}) & = \exp \left( \frac{\eta}{2\gamma c} \left( \beta + \frac{c}{\eta} \right)^{2} \right) \frac{S_{\infty}}{S_{0}} \left( \frac{\eta}{\gamma c}  \beta (1-S_{0}) + \frac{1}{\gamma}  + \frac{\frac{\partial S_{\infty}}{\partial \beta}}{S_{\infty}} \right) \\
& + 2 \beta S_{\infty} \sqrt{\frac{(-\eta)}{2\gamma c}} \int_{\sqrt{\frac{-\eta}{2\gamma c}} \left( \beta S_{\infty} + \frac{c}{\eta}\right)}^{\sqrt{\frac{-\eta}{2\gamma c}} \left( \beta + \frac{c}{\eta}\right)} e^{-v^2} dv \left( \frac{1}{\beta} + \frac{\frac{\partial S_{\infty}}{\partial \beta}}{S_{\infty}} \right) \\
& + \beta S_{\infty} \frac{\eta}{\gamma c} \left( S_{\infty} + \beta \frac{\partial S_{\infty}}{\partial \beta} \right) \exp \left( \frac{\eta}{2\gamma c} \left( \beta S_{\infty} + \frac{c}{\eta} \right)^{2} \right) .
\end{align*}
Equating the derivatives of the left-hand and right-hand sides and using Expression (\ref{eq:S_infty_naive_reformulate}) and rearranging yield  
\begin{align*}
& \exp \left( \frac{\eta}{2\gamma c} \left( \beta S_{\infty} + \frac{c}{\eta} \right)^{2} \right) \left( \left( \frac{\beta}{\gamma} - \frac{1}{S_{\infty}} \right) \frac{\partial S_{\infty}}{\partial \beta} + \frac{1}{\gamma} \left( S_{\infty} - \frac{\gamma}{\beta} \right) \right) \\
= & \exp \left( \frac{\eta}{2\gamma c} \left( \beta + \frac{c}{\eta} \right)^{2} \right) \frac{S_{\infty}}{S_{0}} \left( \frac{1}{\gamma} \left( 1+ \frac{\eta \beta (1-S_{0})}{c}  - \frac{\gamma}{\beta} \right) \right) .
\end{align*}
Thus, 
\begin{align}\label{eq:S_infty_gamma_beta}
\left( \frac{\beta}{\gamma} - \frac{1}{S_{\infty}} \right) \frac{\partial S_{\infty}}{\partial \beta} & = \frac{\exp \left( \frac{\eta}{2\gamma c} \left( \beta + \frac{c}{\eta} \right)^{2} \right)}{\exp \left( \frac{\eta}{2\gamma c} \left( \beta S_{\infty} + \frac{c}{\eta} \right)^{2} \right)} \frac{S_{\infty}}{S_{0}} \frac{1}{\gamma} \left( \varepsilon(0) - \frac{\gamma}{\beta} \right) -  \frac{1}{\gamma} \left( S_{\infty} - \frac{\gamma}{\beta} \right) .
\end{align}
Since $S_{\infty}<\frac{\gamma}{\beta}$ follows from Proposition \ref{prop:epidemic size},\footnote{In fact, Equation (\ref{eq:S_infty_gamma_beta}) itself yields $S_{\infty} \neq \frac{\gamma}{\beta}$. Since $\dot{I}(0) \geq 0$, we have $\varepsilon(0) \geq \frac{\gamma}{\beta S_{0}} > \frac{\gamma}{\beta}$. Since the first-term of the right-hand side of (\ref{eq:S_infty_gamma_beta}) is not zero, it cannot be the case that $S_{\infty}=\frac{\gamma}{\beta}$.} it follows that 
\begin{align*}
\frac{\partial S_{\infty}}{\partial \beta} & = \frac{S_{\infty}}{\beta}\left(\frac{\exp \left( \frac{\eta}{2\gamma c} \left( \beta + \frac{c}{\eta} \right)^{2} \right)}{\exp \left( \frac{\eta}{2\gamma c} \left( \beta S_{\infty} + \frac{c}{\eta} \right)^{2} \right)} \frac{S_{\infty}}{S_{0}} \frac{\varepsilon(0) - \frac{\gamma}{\beta}}{S_{\infty}-\frac{\gamma}{\beta}} -  1\right) <0.
\end{align*}
\newline

\noindent \text{Case 3.} The case with $\beta \in [\overline{\beta}, \frac{c}{(-\eta)I_{0}}]$ is analogous to Case 1, and thus the proof is omitted. 
\end{proof}

\begin{rem}\label{rem:pless1}
Recall that we have assumed $p_{i}(t) <1$ in deriving equation (\ref{eq:optimal_distancing}). We show that this is indeed the case in three steps. First, the proof of the inequality $S_{0} e^{- \frac{\beta}{\gamma}} \leq S_{\infty}$ in Proposition \ref{prop:epidemic size} holds for any SIR dynamics (\ref{eq:S_dot}), (\ref{eq:I_dot}) and (\ref{eq:R_dot}) with $\varepsilon(\cdot) \in [0,1]$. Especially, it holds for the model with the endogenous cost of infection in which $\eta$ evolves according to (\ref{eq:eta_dot}). Second, $\frac{1-p_{\infty}}{1-p(0)} = \frac{S_{\infty}}{S_{0}}>0$ holds, where the equality follows from observations in the proof of Lemma \ref{lemma:eta} and the inequality from the first step. Third, $p_{i}$, which follows (\ref{eq:p_dot}), is weakly increasing and satisfies  $p=p_{i}$ in equilibrium. Then, $p(t) \leq p_{\infty}:=\displaystyle \lim_{t \rightarrow \infty} p(t) < 1$, as desired.
\end{rem}

\begin{proof}[{\bf Proof of Lemma \ref{lemma:eta}}]
We prove (\ref{eq:eta_i_solved}) in two steps. First, it follows from (\ref{eq:p_dot}) that 
\begin{equation*}
\frac{d}{dt} \log (1-p_{i}(t)) = - \frac{\dot{p}_{i}(t)}{1-p_{i}(t)} = - \varepsilon_{i} (t) \beta I(t).
\end{equation*}
Integrating both sides from some $t_0$ to $t_1> t_0$ and taking the exponential yield
\begin{equation}\label{eq:p_ratio}
\frac{1-p_{i}(t_{1})}{1-p_{i}(t_{0})} = \exp \left( - \int_{t_{0}}^{t_{1}} \beta \varepsilon_{i} (t) I(t) dt \right).
\end{equation}

Second, since (\ref{eq:eta_dot}) is a linear first-order differential equation, let $$\mu(t) := e^{-\rho t} \exp \left( - \beta \int_{0}^{t} \varepsilon_{i}(\tau)I(\tau) d\tau \right)$$ be the integrating factor. Since $\frac{d}{dt} \left[ \mu(t) \eta_{i}(t) \right] = \mu (t) \left( \dot{\eta}_{i}(t) - (\rho + \beta \varepsilon_{i}(t)I(t)) \eta_{i}(t) \right)$, it follows that
\begin{align*}
& \frac{d}{dt} \left[ \mu(t) \eta_{i}(t) \right] = \mu(t) \left( (\pi_{S} - \rho V_{I}) -  \frac{c}{2} (1-\varepsilon_{i}(t))^{2} \right).
\end{align*}
Integrating both sides on $[t, \infty)$ and using the transversality condition give
\begin{align*}
& e^{-\rho t}\exp \left( - \beta \int_{0}^{t} \varepsilon_{i}(\tau)I(\tau) d\tau \right) \eta_{i}(t) \\
& = \int_{t}^{\infty} e^{-\rho s} \exp \left( - \beta \int_{0}^{s} \varepsilon_{i}(\tau)I(\tau) d\tau \right) \left( (\pi_{S} - \rho V_{I}) -  \frac{c}{2} (1-\varepsilon_{i}(s))^{2} \right) ds.
\end{align*}
Thus, 
\begin{align}
\eta_{i}(t) & = - \int_{t}^{\infty} e^{-\rho (s-t)} \frac{\exp \left( - \beta \int_{0}^{s} \varepsilon_{i}(\tau)I(\tau) d\tau \right)}{\exp \left( - \beta \int_{0}^{t} \varepsilon_{i}(\tau)I(\tau) d\tau \right)} \left( (\pi_{S} - \rho V_{I}) -  \frac{c}{2} (1-\varepsilon_{i}(s))^{2} \right) ds \label{eq:eta_i_derivation} \\
& = - \int_{t}^{\infty} e^{-\rho (s-t)} \frac{1-p_{i}(s)}{1-p_{i}(t)} \left( (\pi_{S} - \rho V_{I}) -  \frac{c}{2} (1-\varepsilon_{i}(s))^{2} \right) ds, \notag
\end{align}
where the last equality used (\ref{eq:p_ratio}).

Next, we show (\ref{eq:eta_solved}) in two steps. First, observe that (\ref{eq:S_dot}) can be rewritten as
\begin{equation*}
\frac{d}{dt} \log (S(t)) = - \beta \varepsilon (t) I(t). 
\end{equation*}
Integrating both sides from some $t_0$ to $t_1> t_0$ and taking the exponential yield
\begin{equation}\label{eq:S_ratio}
\frac{S(t_{1})}{S(t_{0})} = \exp \left( - \int_{t_{0}}^{t_{1}} \beta (t) \varepsilon (t) I(t) dt \right).
\end{equation}

Second, in an equilibrium, (\ref{eq:eta_i_derivation}) reduces to 
\begin{align*}
\eta(t) & = - \int_{t}^{\infty} e^{-\rho (s-t)} \frac{\exp \left( - \beta \int_{0}^{s} \varepsilon(\tau)I(\tau) d\tau \right)}{\exp \left( - \beta \int_{0}^{t} \varepsilon(\tau)I(\tau) d\tau \right)} \left( (\pi_{S} - \rho V_{I}) -  \frac{c}{2} (1-\varepsilon(s))^{2} \right) ds \\
& = - \int_{t}^{\infty} e^{-\rho (s-t)} \frac{S(s)}{S(t)} \left( (\pi_{S} - \rho V_{I}) -  \frac{c}{2} (1-\varepsilon(s))^{2} \right) ds, 
\end{align*}
where the last equality used (\ref{eq:S_ratio}).
\end{proof}

\begin{proof}[{\bf Proof of Lemma \ref{lemma:eta_bounds}}]
We first show (\ref{eq:eta_infty}). We rearrange (\ref{eq:eta_solved}) as 
\begin{align}\label{eq:eta_rewritten_infty}
\eta(t) = \int_{t}^{\infty} e^{-\rho (s-t)} \frac{S(s)}{S(t)} (\rho V_{I} - \pi_{S}) ds +  \int_{t}^{\infty} e^{-\rho (s-t)} \frac{S(s)}{S(t)} \frac{c}{2} (1-\varepsilon(s))^{2}  ds .
\end{align}
For the first term of (\ref{eq:eta_rewritten_infty}), since $\rho V_{I}-\pi_{S} <0$, 
\begin{align*}
\frac{\rho V_{I} - \pi_{S}}{\rho} = \int_{t}^{\infty} e^{-\rho (s-t)} (\rho V_{I} - \pi_{S}) ds \leq \int_{t}^{\infty} e^{-\rho (s-t)} \frac{S(s)}{S(t)} (\rho V_{I} - \pi_{S}) ds & \leq  \frac{S(\infty)}{S(t)} \frac{\rho V_{I} - \pi_{S}}{\rho}.
\end{align*}
As $t \rightarrow \infty$, the first term of (\ref{eq:eta_rewritten_infty}) converges to $- \frac{\pi_{S}-\rho V_{I}}{\rho}$. For the second term of (\ref{eq:eta_rewritten_infty}), observe $\displaystyle I_{\infty}:= \lim_{t \rightarrow \infty} I(t)=0$. This is because, if $I_{\infty}>0$, then $R$ is unbounded, which is impossible. By optimality condition (\ref{eq:optimal_distancing}), $\displaystyle \lim_{t \rightarrow \infty} \varepsilon_{i}(t) =1$. Then, for any small number $\kappa>0$, there exists $t_{0} \in [0, \infty)$ such that if $t \geq t_{0}$ then 
\begin{equation*}
0 \leq \int_{t}^{\infty} e^{-\rho (s-t)} \frac{S(s)}{S(t)} \frac{c}{2} (1-\varepsilon(s))^{2}  ds  \leq \int_{t}^{\infty} e^{-\rho (s-t)} \frac{c}{2} (1-\varepsilon(s))^{2}  ds \leq \frac{c \kappa^{2}}{2 \rho}.
\end{equation*}
Thus, 
\begin{equation*}
0 \leq \lim_{t \rightarrow \infty} \int_{t}^{\infty} e^{-\rho (s-t)} \frac{S(s)}{S(t)} \frac{c}{2} (1-\varepsilon(s))^{2}  ds \leq \frac{c\kappa^{2}}{2 \rho} .
\end{equation*}
Since $\kappa$ is arbitrary, the second term of (\ref{eq:eta_rewritten_infty}) converges to zero. Hence, we obtain (\ref{eq:eta_infty}), as desired. 

As for the bounds, the lower bound is obtained by replacing $\varepsilon(t)=1$ and $\frac{S(s)}{S(t)} =1$ for all $s \geq t$ in (\ref{eq:eta_solved}). For the upper bound, it follows from (\ref{eq:eta_dot}) that $\dot{\eta}_{i}(t) >0$ if and only if 
\begin{align*}
\eta(t) > - \frac{\pi_{S} - \rho V_{I} - \frac{c}{2}(1-\varepsilon(t))^{2}}{\rho + \varepsilon(t) \beta I(t)}.
\end{align*}
If $\eta(t)$ satisfies $\eta(t) > - \frac{\pi_{S} - \rho V_{I} - \frac{c}{2}}{\rho + \beta}$, then 
from time $t$ on $\eta$ is always increasing, which contradicts the statement that $\eta$ converges to its lower bound as time goes to infinity. 

Next, assume $\dot{\eta}(0)>0$. Observe that $\eta$ is bounded because it is continuous and converges to the finite lower bound (\ref{eq:eta_infty}). Letting $t_{\eta}$ be a time at which $\eta$ attains a maximum, it follows from the assumption $\dot{\eta}(0)>0$ that $\dot{\eta}(t_{\eta})=0$. Thus, 
\begin{align*}
\eta(t_{\eta}) = - \frac{\pi_{S} - \rho V_{I} - \frac{c}{2} (1-\varepsilon(t_{\eta}))^{2}}{\rho + \varepsilon(t_{\eta}) \beta I(t_{\eta})}.
\end{align*}
Substituting for $\beta I(t)$ from equation (\ref{eq:optimal_distancing}) for optimal distancing and rearranging yield
\begin{align}
\eta (t_{\eta}) & =  - \frac{\pi_{S} - \rho V_{I}}{\rho} + \frac{c}{2 \rho} (1-\varepsilon^{2}(t_{\eta})) \label{eq:eta_max} \\
& \leq - \frac{\pi_{S} - \rho V_{I} - \frac{c}{2}}{\rho} . \notag
\end{align}

Finally, we show that the upper bound $- \frac{\pi_{S}- \rho V_{I}+\frac{c}{2}}{\rho}$ is approximately tight when $\dot{\eta}(0)>0$. Substituting (\ref{eq:eta_max}) into optimality condition (\ref{eq:optimal_distancing}) yields the following quadratic equation with respect to $\varepsilon(t_{\eta})$: 
\begin{align*}
\frac{\beta I(t_{\eta})}{2 \rho} \varepsilon^{2}(t_{\eta}) + \varepsilon (t_{\eta}) - 1 + \frac{\beta}{c} I(t_{\eta}) \frac{\pi_{S}-\rho V_{I}-\frac{c}{2}}{\rho} =0.
\end{align*}
This quadratic equation admits a unique solution $\varepsilon (t_{\eta}) \in [0,1]$: 
\begin{align*}
\varepsilon (t_{\eta}) & = - \frac{\rho}{\beta I(t_{\eta})} \left( 1 - \sqrt{1+2\frac{\beta I(t_{\eta})}{\rho} \left( 1-\frac{\beta}{c} I(t_{\eta}) \frac{\pi_{S}-\rho V_{I}-\frac{c}{2}}{\rho} \right) } \right).
\end{align*}
Since $1-\sqrt{1+2x} \approx -x$ and $1-\sqrt{1+2x} \geq -x$, 
\begin{align*}
\varepsilon (t_{\eta}) & \approx  \frac{\rho}{\beta I(t_{\eta})} \left( \frac{\beta I(t_{\eta})}{\rho} \left( 1-\frac{\beta}{c} I(t_{\eta}) \frac{\pi_{S}-\rho V_{I}-\frac{c}{2}}{\rho} \right) \right) =  1-\frac{\beta}{c} I(t_{\eta}) \frac{\pi_{S}-\rho V_{I}-\frac{c}{2}}{\rho} .
\end{align*}
Comparing the last equation with optimality condition (\ref{eq:optimal_distancing}), we obtain
$\eta(t_{\eta}) \approx - \frac{\pi_{S}-\rho V_{I}-\frac{c}{2}}{\rho}$.
\end{proof}

\begin{proof}[{\bf Proof of Proposition \ref{prop:phase_comparison_full}}]
Recall that 
$$\frac{dI}{dS} = -1 + \frac{\gamma}{\beta S \max  \left ( 0, 1+ \frac{\beta \eta}{c}I\right) },$$
and denote the solution path of the model with endogenous cost of infection by $(S_e,I_e)$ and its co-state by $\eta_e$.
 By construction, $\eta_e(\cdot) \in[\eta_L,\eta_H]$. Therefore, for any fixed values of $S$ and $I$, the following chain of inequalities obtains: $\frac{d I_H}{d S_H} \leq  \frac{d I_e}{d S_e} \leq  \frac{d I_L}{d S_L}$. Finally, recall that all three solution paths go through $(S_0,I_0)$.
 
Consider first the solution path of the model with an endogenous cost of infection and the model with the fixed cost of infection $\eta_L$. Since $\frac{dI_{L}}{dS_{L}} \geq \frac{dI_{e}}{dS_{e}}$, at any point of intersection the solution path of the model with the fixed cost $\eta_L$ intersects the model with the endogenous cost from below. Hence, for $\delta >0$ small enough  $I_L(S_0-\delta)\leq I(S_0-\delta)$. But then there can be no intersection for any $S<S_0$ as otherwise at such an intersection $\frac{dI_{L}}{dS_{L}} <\frac{d I_e}{d S_e}$. Thus, $I_L(S)\leq I_e(S)$.

The proof for the case with the fixed cost of infection $\eta_H$ is analogous and, therefore, $I_H(S)\geq I_e(S)$. 
\end{proof}

\section{Parameters and Computational Algorithm}
\label{app:parameters}

We simulate the model at a daily frequency. We follow \citet{Farboodi_et_al_20} for most model parameters as summarized in Table \ref{table:parameters}. We set $\gamma = 1/7$, assuming that the average length of disease is $7$ days. For the transmission rate $\beta$ for the baseline simulation of the endogenous cost of infection model, we assume that the initial growth rate $\frac{\dot{I}(0)}{I(0)}$ without behavior is $0.3$. Since it is given as $\beta-\gamma$ for the dynamics of the standard SIR model with $S_{0}=1$, we set $\beta = 0.3+\gamma = 0.443$. This gives $R_{0} =  3.1$ without behavior. We vary $\beta$ for various numerical simulations. For $I_{0}$, we match 194 people who died from COVID-19 in the US on or before March 18, a week after the pandemic declaration of the WHO on March 11, 2020. Given a population of 328 million and an IFR of $0.0062$, we set $I_{0} = 0.95 \times 10^{-4}$. We take $\rho=\tilde{\rho} + \lambda= (0.05+0.67)/365$, where $\rho$ captures a 5 percent annual discount rate, and $\lambda$ implies an expected time until the arrival of a cure of 1.5 years as in \citet{alvarez2020simple} and \citet{Farboodi_et_al_20}.

For the flow payoff, we normalize it to be $-(1-\varepsilon(t))^{2}$. Thus, we set $c=2$ and $\pi_{S} =0$. To compute the parameter $\eta$ of the constant cost of infection model, we follow the same steps as in \citet{Farboodi_et_al_20}. We assume the value of a statistical year of life to be US \$ $270,000$ and an average remaining life expectancy of COVID-19 victims to be 14.5 years, which gives US\$ $3,915,000$ where the numerical values are taken from \citet{Hall_Jones_Klenow_20}. Hence, to avoid a 0.1 percent probability of death an individual would be willing to pay US\$ $0.001 \times 3,915,000$. Using the discount rate to translate this into flow units we obtain US\$ $\rho \cdot 3,915$ as the willingness to pay to avoid the 0.1 percent probability of death. To translate this into utils, we also use the US per capita consumption from \citet{Hall_Jones_Klenow_20} of US\$ 45,000 per year implying that an individual is willing to give up $\frac{3,915 \rho \cdot 365}{45,000}  =31.755 \rho$ in terms of annual consumption units, i.e., $\varepsilon=1-31.755\rho$, to avoid a 0.1 percent risk of death. Applying the assumed utility function, an individual, who is willing to give up 31.755 $\rho$ units of consumption per period to avoid a 0.1 percent risk of death, is indifferent between this and full exposure with a 0.001 risk of death which has a utility cost of $v$:
\begin{align*}
-\frac{(1-1)^2}{\rho}-0.001v = -\frac{(1-31.755 \rho)^2}{\rho}.
\end{align*}
Multiplying this value of life in utils by the death rate of 0.0062 \citep[also from][]{Hall_Jones_Klenow_20} yields a cost of infection $\eta =-2761.63$.

For the endogenous cost of infection model, we set $\pi_R=0$ and $\pi_{I} = -399.96$ so that $V_{I} = \frac{\pi_{I}}{\rho+\gamma} =\eta$ works as the lower bound of $\eta(t)$ in the endogenous cost of infection model. The upper bound of $\eta$ is $-2761.63 + \frac{c/2}{\rho} = -2254.68$, which we also use in the constant cost of infection model.

\begin{table}[]\begin{center}
	\caption{\emph{Table of Baseline Parameters for Numerical Analysis.}}\label{table:parameters}
	{\footnotesize
\begin{tabular}{cccc}
\textbf{Parameter} & \textbf{Description}        & \textbf{Value}          & \textbf{Source}                                                                                \\ \hline
$\gamma$           & Recovery Rate               & $1/7$           & \citet{Farboodi_et_al_20}                                                                    \\
$\beta$            & Transmission Rate           & $0.3+\gamma$            & \citet{Farboodi_et_al_20}                                                                         \\
$I_0$              & Initial Seed of Infections  & $0.95\times 10^{-4}$    & \begin{tabular}[c]{@{}c@{}}Based on death toll in the \\ US before March 19, 2020\end{tabular} \\
$\tilde{\rho}$             & Discount Rate               & $0.05/365$      & \citet{Farboodi_et_al_20}                                                                       \\
$\lambda$          & Arrival Rate of Cure        & $0.67/365$      & \citet{Farboodi_et_al_20}                                                                      \\
$c$                & Cost of Distancing          & $2$                       & Normalization                                                                                  \\
$\pi_S$            & Flow Payoff of Susceptibles & $0$                       & Normalization                                                                                  \\ 
$\eta$             & Cost of Infection           & $\{-2761.63,-2254.68\}$ & \citet{Hall_Jones_Klenow_20}      \\ \hline                                                         
\end{tabular}}
	\end{center}
\end{table}

We have solved the constant cost of infection model using the fourth-order Runge-Kutta method. For the endogenous cost of infection model, recall that the equilibrium of the model is characterized as follows. First, $(S,I,R)$ follow (\ref{eq:S_dot}), (\ref{eq:I_dot}), and (\ref{eq:R_dot}) with the initial condition $(S(0),I(0),R(0)) = (S_{0},I_{0},0)$, where $\varepsilon(t)=1+\frac{\beta \eta(t)}{c} I(t)$ is the average exposure. Second, $\eta$ follows equation (\ref{eq:eta_dot}) with $\displaystyle \lim_{t \rightarrow \infty} \eta(t)=- \frac{\pi_{S}-\rho V_{I}}{\rho}$ as in (\ref{eq:eta_infty}). 

To numerically solve $(S,I,R,\eta)$, we set $\eta(T)=-\frac{\pi_{S}-\rho V_{I}}{\rho}$ at $T=400\times 365$ (days). Then, given $\eta$, we solve for $(S,I,R)$ with the initial condition. In turn, given $(S,I,R)$, we solve for $\eta$ with the terminal condition $\eta(T)=-\frac{\pi_{S}-\rho V_{I}}{\rho}$. We iterate the procedure until the sum of the distances of $(S,I,R,\eta)$ in two successive iterations is below a threshold value. To facilitate the computation, at each iteration, when $S(t)-S(t+1)$ and $I(t+1)$ are below threshold values, we have terminated the simulation of $(S,I,R)$ at $t+1$, and we start the computation of $\eta$ with $\eta(t+1)=-\frac{\pi_{S}-\rho V_{I}}{\rho}$ and $(S,I,R)$. Once the iterations end, we have checked whether $\varepsilon (\tau) \in [0,1]$ for every time $\tau$. The right panel of Figure \ref{fig:endogenous_cost} depicts the peak prevalence when $\varepsilon(\tau) \in [0,1]$ for every time $\tau$.

\end{document}